\documentclass[aps,prd,preprint,superscriptaddress,amsfonts,amssymb,amsmath,nofootinbib]{revtex4-1}
\usepackage{graphicx, bm, color, ulem, latexsym, float}
\usepackage[colorlinks, linkcolor=magenta, anchorcolor=blue, citecolor=blue]{hyperref}
\usepackage{multirow,subfigure}

\begin{document}
	
\title{Stimulated radiation from superradiant scalar cloud in scalar-tensor theory}



\author{Wenyi Wang}
\email{wangwy@mails.ccnu.edu.cn}
\affiliation{Institute of Astrophysics, Central China Normal University, Wuhan 430079, China}

\author{Sousuke Noda}
\email{snoda@cc.miyakonojo-nct.ac.jp}
\affiliation{National Institute of Technology, Miyakonojo College, Miyakonojo 885-8567, Japan}

\author{Taishi~Katsuragawa}
\email{taishi@ccnu.edu.cn}
\affiliation{Institute of Astrophysics, Central China Normal University, Wuhan 430079, China}
\affiliation{Laboratory for Compact Object Astrophysics and Astronomical Technology, Central China Normal University, Wuhan 430079, China}

\begin{abstract}

Scalar–tensor theories predict fundamental scalar fields of considerable interest in astrophysics and cosmology.
We investigate the superradiant instability of scalar clouds around Kerr black holes, showing that stimulated decay generates detectable electromagnetic signals.
The growth of the superradiant scalar cloud differs from that of other bosonic fields and depends sensitively on the matter distribution surrounding the black hole, which originates from the scalar-matter coupling realized by the chameleon mechanism in modified gravity theories.
In non-uniform matter distributions, stimulated emission from scalar clouds offers an observational signature that distinguishes fundamental scalars from other light bosonic fields.
\end{abstract}

\maketitle

\section{Introduction}


Many classes of modified gravity theories predict the existence of new fundamental scalar fields~\cite{Sotiriou:2008rp, Nojiri:2010wj, Clifton:2011jh, Nojiri:2017ncd}.
These scalar fields arise as additional scalar degrees of freedom, denoted by $\varphi$ in scalar–tensor theories, and have been extensively explored in astrophysical and cosmological contexts, including as the inflaton~\cite{linde1982new,linde1983chaotic}, dynamical dark energy~\cite{ratra1988cosmological,chen2022towards}, and, in certain scenarios, as potential dark matter candidates~\cite{Cembranos:2008gj, Choudhury:2015zlc, Aoki:2016zgp, Babichev:2016hir, Katsuragawa:2016yir, Katsuragawa:2017wge, Yadav:2018llv, Burrage:2018zuj, Shtanov:2021uif, Kading:2023hdb, Shtanov:2024nmf, Shtanov:2025nue}.
A distinctive property of scalars arising in modified gravity theories is their environmental dependence, known as the chameleon mechanism, through which the coupling to matter modifies their effective mass~\cite{Brax:2013ida, Burrage:2017qrf, Brax:2021wcv, Dong:2013swa, Solomon:2022qqf}.
This property distinguishes such scalars from other bosonic fields.


Moreover, scalars can interact with the electromagnetic field through the quantum trace anomaly, which induces a coupling to two photons.
The relevant tree-level interaction between the scalar and photons can be written as~\cite{Katsuragawa:2016yir}
\begin{align}
\label{scalar-photon}
    \mathcal{L}_{\varphi \gamma \gamma} 
    = 
    \frac{1}{4} \frac{\beta_{\gamma}}{M_{\mathrm{pl}}} \varphi F_{\mu\nu}F^{\mu\nu}
    \, ,
\end{align}
where $M_{\mathrm{pl}}$ is the reduced Planck mass, $\beta_{\gamma}$ is the coupling constant between the scalar and the electromagnetic field, $\varphi$ is a scalar field, and $F_{\mu\nu}$ is the field-strength tensor.
This interaction yields the scalar-field lifetime decaying to photons~\cite{Katsuragawa:2021wmw, Shtanov:2021uif}
\begin{align}
\label{scalar-lifetime}
    \tau_{\varphi} 
    \simeq 
    3 \times 10^{39} \cdot \left(\frac{\mu_{0}}{\mathrm{eV}}\right)^{-3}\cdot \beta_{\gamma}^{-2}~\mathrm{s}
    \, ,
\end{align}
where $\mu_{0}$ is the scalar mass.  
Various ground-based experiments have constrained the relevant parameter space~\cite{ADMX:2010ndb, Steffen:2010ze, CAST:2018bce, Homma:2022ktv, Wang:2024gmn, Homma:2024aan, Hasada:2025rek}; for instance, collider experiments place the bound $\beta_{\gamma} \lesssim 2\times 10^{15}$ for $10^{-11}~\mathrm{eV} \lesssim \mu \lesssim 10^{4} ~\mathrm{eV}$~\cite{Brax:2009aw}.


Light bosonic fields, such as axions, can generate superradiant clouds around rotating black holes by extracting their rotational energy~\cite{Cardoso:2005vk, Dolan:2007mj, Brito:2015oca, Spieksma:2023vwl}, a phenomenon known as the superradiant instability.
Previous studies have shown that superradiant axion clouds can produce observable electromagnetic signals through stimulated decay into photons.
In addition to spontaneous decay into two photons, there exists an induced channel in which ambient photons stimulate the axion to decay into photons.
This process is analogous to the inverse Primakoff effect~\cite{Sikivie:1983ip, Dent:2020jhf}, where an axion is converted into a photon in the presence of an external magnetic field, and existing experiments for axions and axion-like particles have incorporated this stimulated decay channel in their analyses~\cite{Homma:2023amn, Nakamura:2025mkp}.
Unlike axions or other bosonic fields, fundamental scalars in scalar–tensor theories can exhibit strong environmental dependence through the chameleon mechanism, implying that the growth of superradiant scalar clouds varies significantly with the surrounding matter distribution~\cite{Cardoso:2013opa, Cardoso:2013fwa, Zhang:2014kna, Nakamura:2020ihr, Lingetti:2022psy}.
As a consequence, the stimulated electromagnetic emission generated by superradiant scalar clouds may differ qualitatively from that associated with other bosonic fields.


Motivated by recent developments in superradiant instabilities and stimulated decays, this work extends the existing framework of superradiant axion clouds and photon emission to fundamental scalars arising in scalar–tensor theories.
We analyze differences in stimulated emission signals from superradiant scalar clouds for various matter distributions around Kerr black holes.
For uniform matter distributions, we study the stimulated emission associated with different scalar modes in non-spherical superradiant clouds, in close analogy with other bosonic fields.
For non-uniform matter distributions, we first examine the superradiant instability of the massive scalar field and then analyze the corresponding features of the stimulated emission from the resulting scalar cloud.
We show that this environmental dependence provides a potential means to distinguish fundamental scalars from other bosonic fields.


The paper is organized as follows.
In Sec.~\ref {sec-ST}, we briefly review the framework of scalar–tensor theories and the perturbation equations for the scalar field.
In Sec.~\ref {sec-uniform matter}, we introduce the evolution equations for scalar and photon number densities in a non-spherical superradiant cloud and present our numerical results for different scalar modes.
In Sec.~\ref {non-uniform}, we discuss the superradiant instability of the scalar field and the associated stimulated emission in the presence of non-uniform matter distributions.
Finally, Sec.~\ref{sec-cd} is devoted to conclusions and discussion.
Throughout this paper, we use natural units, $G = \hbar = c = 1$.

\section{Scalar-tensor theory and field equations}
\label{sec-ST}


We consider the following action of a scalar–tensor theory:
\begin{align}
\label{action}
    S
    =
    \int d^{4} x \sqrt{-g} \left(\frac{R}{16\pi } - \frac{1}{2} g^{\mu\nu} \partial_{\mu}\Phi \partial_{\nu} \Phi - \frac{V(\Phi)}{16\pi }\right) + S(\Psi_{\text{m}}; A(\Phi)^{2} g_{\mu\nu})
    \, ,
\end{align} 
where $R$ is the Ricci scalar, $\Phi$ is the scalar field, which is coupled to the metric $A(\Phi)^{2} g_{\mu\nu}$, and $\Psi_{\text{m}}$ denotes the matter fields.
Note that even if a nonminimal coupling of the form $f(\Phi) R$ is included, it can be eliminated by a conformal transformation of the metric, $g_{\mu\nu}\to f(\Phi)g_{\mu\nu}$, leading to an action equivalent to Eq.~\eqref{action}.
Following Refs.~\cite{Cardoso:2013opa, Cardoso:2013fwa, Lingetti:2022psy}, we assume that both the potential $V(\Phi)$ and the coupling function $A(\Phi)$ are generic and analytic in the vicinity of $\Phi \sim \Phi^{(0)}$, and can be expanded as
\begin{align}
    V(\Phi)
    &=
    \sum_{n=0} V_{n}(\Phi-\Phi^{(0)})^{n}
    \, , \\
    A(\Phi)
    &=
    \sum_{n=0} A_{n}(\Phi-\Phi^{(0)})^{n}
    \, .
\end{align}
Here, $V_n$ and $A_n$ are the expansion coefficients, and $\Phi^{(0)} = \text{const}$ is a solution to the field equations.


To study the superradiant instability in scalar–tensor theories, we work in the Kerr spacetime, which provides a stationary and axisymmetric background.
Expanding the field equations derived from the action~\eqref{action} to first order in the perturbation $\varphi \equiv \Phi-\Phi^{(0)} \ll 1$, we obtain the Klein–Gordon equation
\begin{align}
\label{KG eq}
    \left[\square - \mu_{\mathrm{eff}}^{2}(r,\theta) \right]\varphi = 0
    \, ,
\end{align}
where the effective mass of the scalar field $\mu_{\mathrm{eff}}$ is given by
\begin{align}
\label{effetive-mass}
    \mu_{\mathrm{eff}}^{2}(r,\theta)
    =
    \frac{V_{2}}{8\pi }
    - \frac{2A_{2}}{A_{0}} T(r,\theta)
    \, .
\end{align}
Here, $T(r,\theta)$ denotes the trace of the energy–momentum tensor of the ambient matter fields around the Kerr black hole other than the scalar field.
We have set $A_{1}=0$ to admit vacuum solutions consistent with general relativity, and $V_{0}=V_{1}=0$ to ensure asymptotic flatness (see Ref.~\cite{Cardoso:2013fwa} for details).
The first term in Eq.~\eqref{effetive-mass} corresponds to the bare mass of the scalar field. 
In contrast, the second term describes the modification of the effective mass induced by the surrounding matter distribution, that is, the chameleon mechanism.


To investigate the chameleon mechanism in the scalar field, we decompose the matter source $T(r,\theta)$ into two components:
\begin{align}
\label{T-decomposition}
    T(r,\theta) 
    =
    T_{0} + T_{1}(r,\theta)
    \, ,
\end{align}
where $T_{0}=const$ describes a uniform matter distribution, while $T_{1}(r,\theta)$ encodes spatial inhomogeneities.
Throughout this work, we further assume that the non-uniform part can be written in the following separable form:
\begin{align}
\label{T1-separable}
    T_{1}(r,\theta) 
    = 
    -\frac{A_{0}}{2A_{2}}\frac{G(r) + F(\theta)}{r^{2}+a^{2} \cos^{2}\theta}
    \, .
\end{align}
The functions $G(r)$ and $F(\theta)$ represent the radial and angular contributions to the effective mass arising from the coupling between the scalar field and the surrounding matter distribution.
With this decomposition, the effective mass squared of the scalar field, defined in Eq.~\eqref{effetive-mass}, can be expressed as~\cite{Cardoso:2013opa, Cardoso:2013fwa}:
\begin{align}\label{Eq: eff_mass_matter}
    \mu_{\mathrm{eff}}^{2}(r,\theta)
    &=
    \mu_{0}^{2} + \frac{G(r) + F(\theta)}{r^{2}+a^{2} \cos^{2}\theta}
    \, .
\end{align}
Here, $a$ denotes the angular momentum per unit mass of the background Kerr spacetime.
$\mu_{0}$ plays the role of the canonical bare mass of the scalar field, which corresponds to the effective mass in the uniform-matter case:
\begin{align}
\label{mu0-def}
    \mu_{0}^2 
    \equiv 
    \frac{V_{2}}{8\pi} - \frac{2A_{2}}{A_{0}} T_{0}
    \, ,
\end{align}
Therefore, for the scalar field in a uniform matter distribution ($T(r,\theta) = T_{0} \neq 0$) or vacuum ($T(r,\theta) = T_{0} = 0$), the effective mass is spatially constant.


The scalar field is expanded in separable modes of the following form~\cite{Cardoso:2013fwa}:
\begin{align}
\label{Eq: phi_decomposition}
    \varphi
    &=
    R_{lm}(r) S_{lm}(\theta) e^{-i\omega t + i m \phi}
    \, ,
\end{align}
where the scalar field $\varphi$ is decomposed into a radial function $R_{lm}(r)$ and an angular function $S_{lm}(\theta)$, corresponding to modes with angular quantum number $l$ and azimuthal quantum number $m$.
The frequency $\omega$ is generally a complex number, $\omega = \omega_{R} + i \omega_{I}$, and the sign of the imaginary part $\omega_{I}$ determines whether the scalar field grows or decays exponentially as $e^{\omega_{I} t}$.
With this ansatz, Eq.~\eqref{KG eq} separates into two ordinary differential equations for the radial and angular functions:
\begin{align}
    &\Delta \frac{d}{dr} \left(\Delta \frac{dR_{lm}}{dr} \right)
    +\left\{
    K 
    - \Delta \left[ G(r)+\omega^{2}a^{2}+\mu_{0}^{2}r^{2} + \lambda_{lm} \right]
    \right \} 
    R_{lm}=0,\\
    &\frac{1}{\sin\theta} \frac{d}{d\theta}\left(\sin\theta \frac{d S_{lm}}{d\theta}\right)
    + \left[a^{2}(\omega^{2}-\mu_{0}^{2}) \cos^{2}\theta - \frac{m^{2}}{\sin^{2}\theta} - F(\theta) +\lambda_{lm}\right]S_{lm}=0.
\end{align}
Here, $K=\omega^{2}(r^{2}+a^{2})^{2} - 4Mam\omega r +m^{2}a^{2}$ and $\Delta = r^{2} - 2Mr + a^{2}$, where $M$ is the mass of the Kerr black hole, and $\lambda_{lm}$ is a separation constant.
In the following analysis, we focus on the case $\mu_{0}^{2} \neq 0$ and set $F(\theta)=0$, which captures the leading environmental effects induced by distributed matter around the Kerr black hole through the chameleon mechanism.
This setup allows us to investigate stimulated emission from superradiant scalar clouds within a physically motivated and analytically tractable framework.

\section{The uniform distribution of matter}
\label{sec-uniform matter}

In this section, we investigate the superradiant instability of the scalar field in scalar-tensor theories in the presence of a uniform matter distribution; that is, $\mu_{0}^{2} \neq 0$, $G(r)=F(\theta)=0$.
We then analyze the stimulated decay of the scalar field within the resulting superradiant cloud.
Based on this analysis, we study the stimulated electromagnetic radiation for two representative modes, namely the $l=m=1$ and $l=2,\,m=1$ modes.

\subsection{Superradiant instability and stimulated decay of scalar cloud}


We begin by briefly reviewing the role of massive bosonic fields in superradiance.
For generic massive bosonic fields, the field mass provides an effective potential barrier at large distances, allowing for the existence of bound states around Kerr black holes.
The superradiant instability arises when these bound-state frequencies satisfy the conditions $\omega_{\text{R}} < m\Omega_\text{H}$ and $\omega_{\text{I}} > 0$, where $\Omega_{\text{H}}$ is the angular velocity of the black hole horizon.
The corresponding growth rate depends sensitively on the dimensionless mass coupling $\mu_{0} M$~\cite{Dolan:2007mj}.
In particular, for $\mu_{0} M \ll 1$, the instability timescale can be analytically approximated, and the corresponding spectrum exhibits a hydrogenic-like structure~\cite{Detweiler:1980uk, Pani:2012bp, Pani:2012vp, Brito:2014wla},
whereas for $\mu_{0} M \gg 1$ the instability is strongly suppressed~\cite{zouros1979instabilities}.


In scalar-tensor theories, however, as shown in Eq.~\eqref{effetive-mass}, scalar fields generically acquire an effective mass that depends on the local matter distribution.
This environmental dependence arises from the chameleon mechanism, in which the scalar field becomes heavy in high-density regions while remaining light in low-density regions.
As a consequence, the effective mass relevant for analyzing the superradiant instability is not a constant but may vary with the surrounding matter configuration.
This feature is intrinsic to scalar-tensor theories and is absent in the standard analyses of superradiant instability within general relativity.


As a first step, in order to make contact with the well-understood case of massive bosonic fields, we consider a uniform matter distribution case $T(r,\theta)=T_0$, for which the effective mass becomes constant.
This setup is also equivalent to the vacuum case for the scalar field with an intrinsic mass in the sense that the mass is a constant. 
Note that the value of the constant mass can be shifted due to the effect of the matter $T_0$ in Eq.~\eqref{mu0-def}.
In this case, the bound-state frequency spectrum in the Kerr background can be numerically computed using the continued fraction method~\cite{Dolan:2007mj}.
In the regime of small mass couplings ($\mu_{0} M \ll 1$), the eigenvalue spectrum admits hydrogenic-like solutions~\cite{Detweiler:1980uk,Pani:2012bp,Pani:2012vp,Brito:2014wla}.
The effects of non-uniform matter distributions and the associated chameleon mechanism will be discussed in Sec.~\ref{non-uniform}.


An essential consequence of the superradiant instability is the formation of a superradiant cloud around the Kerr black hole.
As superradiance extracts rotational energy from the black hole, the bound-state modes of the scalar field can undergo exponential growth.
This process leads to the accumulation of scalar particles, resulting in a high-density superradiant scalar cloud surrounding the black hole.
In scalar-tensor theories, as shown in Eq.~\eqref{scalar-photon}, the interaction between the scalar and electromagnetic fields allows the scalar to decay into photons.
In the presence of a superradiant scalar cloud with a large occupation number, this decay process is significantly enhanced by stimulated emission.
As a result, the dense scalar environment generated by superradiance provides a natural setting for stimulated scalar decay into photons.

Motivated by these considerations, we investigate the stimulated photon emission induced by scalar decay within a superradiant cloud.
To this end, we consider the Boltzmann equation governing the time evolution of the photon number density due to scalar decay and inverse decay processes~\cite{Kephart:1994uy, Chen:2020yvx, Chen:2023bne}:
\begin{align}
\label{boltzmann-eq}
\begin{split}
    \frac{d n_{\lambda}(\bf{k})}{dt}
    &=
    \int dX_{\text{LIPS}}^{(3)} 
    \left \{
        f_{\varphi}(\mathbf{p})
        \left[ 1 + f_{\lambda}(\mathbf{k}) \right]
        \left[ 1 + f_{\lambda}(\mathbf{k}^{\prime}) \right]
    \right.
    \\
    & \qquad \qquad\qquad\qquad 
    \left. 
        - f_{\lambda}(\mathbf{ k}) 
        f_{\lambda}(\mathbf{k}^{\prime}) 
        \left[ 1 + f_{\varphi}(\mathbf{p}) \right]
    \right \} 
    \left| M(\varphi \to \gamma\gamma ) \right|^{2}
    \, ,
\end{split}
\end{align}
where the Lorentz-invariant three-body phase space is given by
\begin{align}
    \int dX_{\text{LIPS}}^{(3)} 
    = \int \frac{d^{3} p}{(2\pi)^{3} 2p^{0}}
    \int \frac{d^{3} k_{1}}{(2\pi)^{3} 2k_{1}^{0}}
     \int \frac{d^{3} k_{2}}{(2\pi)^{3} 2k_{2}^{0}}
     (2\pi)^{4} \delta^{(4)}(p-k_{1}-k_{2})
     \, .
\end{align}
The Eq.~\eqref{boltzmann-eq} describes the rate of change of the photon number density $n_{\lambda}$ for helicities $\lambda = \pm 1$, where $f_{i}$ denote the phase-space distribution functions of each species, with $n_{i} = \int d^{3} k_{i}/(2\pi)^{3} f_{i}$.
$M(\varphi \to \gamma\gamma )$ denotes the amplitude for scalar decay into a photon pair.
The scalar and photon occupation numbers are expanded in spherical harmonics $Y_{lm}(\Omega)$ as
\begin{align}
\label{spherical-harmonics}
\begin{split}
    f_{\varphi}(p,r,\Omega,t) 
    &=
    \sum_{lm} f_{\varphi l m} Y_{lm}(\Omega) \Theta(R-r) \Theta(p_{\text{max}}-p) 
    \, ,\\
    f_{\lambda}(k,r,\Omega,t) 
    &=
    \sum_{lm} f_{\lambda l m} Y_{lm}(\Omega)\Theta(R-r) \Theta(k_{+}-k)\Theta(k-k_{-})
    \, .
\end{split}
\end{align}
Here, $R$ denotes the radius of the cloud, $p_{\text{max}}=m_{\varphi} \beta_{\varphi}$ is the maximum scalar momentum inside the cloud, and $\beta_{\varphi}$ is expected to be of the order of the escape velocity.
The quantities $k_{+}$ and $k_{-}$ denote the maximum and minimum photon momenta, respectively.

Using Eqs.~\eqref{boltzmann-eq}--\eqref{spherical-harmonics}, we obtain the evolution equations for the scalar and photon number densities in each $(l,m)$ mode:
\begin{align}
 \label{eq-density-scalar}
    \frac{d n_{\varphi l m}}{dt} 
    &= 
    -\Gamma_{\varphi} \left[n_{\varphi lm} + \frac{8\pi^{2}}{\beta_{\varphi} m_{\varphi}^{3}} E_{lm} - \frac{8\pi^{2}\beta_{\varphi}}{3 m_{\varphi}^{3}} F_{lm} \right]
    \, , \\
\label{eq-density-photon}
    \frac{dn_{\gamma lm}}{dt} 
    &= 
    \Gamma_{\varphi} \left[2 n_{\gamma lm} + \frac{16 \pi^{2}}{\beta_{\varphi} m_{\varphi}^{3}}E_{lm} -\frac{16\pi^{2}}{3m_{\varphi}^{3}}(\beta_{\varphi}+\frac{3}{2})F_{lm}\right] - \Gamma_{e} n_{\gamma lm}
    \, .
\end{align}
Here, $\Gamma_{\varphi} = \tau_{\varphi}^{-1}$ is the scalar decay rate, $m_{\varphi}=\mu_{0}$ is the scalar mass, and $\Gamma_{e}$ denotes the photon escape rate from the cloud.
The coefficients $E_{lm}$ and $F_{lm}$ arise from the scalar--photon interaction terms and are defined as~\cite{Chen:2020yvx, Chen:2023bne}
\begin{align}
    \label{eq-E}
    n_{\varphi }n_{\gamma} 
    &= \sum_{l'm'l''m''} n_{\varphi l'm'} n_{\gamma l'' m''} Y_{l'm'} Y_{l''m''} 
    = \sum_{lm} E_{lm}Y_{lm}
    \, ,\\
    \label{eq-F}
    n_{\gamma}^{2} 
    &= \sum_{l'm'l''m''} n_{\gamma l'm'} n_{\gamma l'' m''} Y_{l'm'} Y_{l''m''} 
    = \sum_{lm} F_{lm}Y_{lm}
    \, .
\end{align}
The quantities $E_{lm}$ and $F_{lm}$ encode the angular-mode coupling between the scalar and photon occupation numbers.
In the following, we discuss the stimulated radiation from superradiant scalar clouds for different scalar modes.

\subsection{$l=m=1$}
\label{uniform-l=1}

The fastest-growing mode is the ``2p'' state ($n=2,\,l=m=1$), as discussed in Refs.~\cite{Chen:2020yvx,Chen:2023bne}.
For the $(l=m=1)$ mode, the scalar wave function is proportional to $Y_{1}^{\pm1}$, which implies that the scalar number density $n_{\varphi}$ is proportional to
$Y_{1}^{\pm1*}Y_{1}^{\pm1} \propto \sin^{2}\theta$, where $\theta$ denotes the polar angle.
Consequently, only the $n_{\varphi 00}$ and $n_{\varphi 20}$ components of the scalar number density are nonzero.

The scalar number density $n_{\varphi}(r,\theta,t)$ can therefore be expressed as
\begin{align}
\begin{split}
    n_{\varphi} 
    &= 
    \Theta(R-r) n_{\varphi}(t) \sin^{2}\theta
    \, , \\
    &= 
    \Theta(R-r) n_{\varphi}(t) 
    \cdot 
    \frac{4\sqrt{\pi}}{3} \left(Y_{00}-\frac{1}{\sqrt{5}}Y_{20}\right)
    \, , \\
    &= 
    \Theta(R-r) [n_{\varphi 00}(t) Y_{00} + n_{\varphi 20} Y_{20}]
    \, ,
\end{split}
\end{align}
which leads to $n_{\varphi 20}(t) = -\sqrt{1/5} n_{\varphi 00}(t)$.
However, to avoid a negative photon density, an additional correction term $n_{\gamma 40}$ must be included alongside the $n_{\gamma 00}$ and $n_{\gamma 20}$ components (see Ref.~\cite{Chen:2023bne} for details).
Accordingly, the total scalar and photon number densities $n_{\varphi}$ and $n_{\gamma}$ are expanded in spherical harmonics as
\begin{align}
\label{eq-components}
\begin{split}
    n_{\varphi} 
    &= 
    n_{\varphi 00}Y_{00} +n_{\varphi 20}Y_{20} + n_{\varphi 40}Y_{40} 
    \,, \\
    n_{\gamma} 
    &= 
    n_{\gamma 00}Y_{00} +n_{\gamma 20}Y_{20} + n_{\gamma 40}Y_{40}
    \, .
\end{split}
\end{align}
The $n_{\varphi 40}$ component is not generated directly by the superradiant instability but arises from the inverse decay of photons.

Using Eqs.~\eqref{eq-density-scalar}, \eqref{eq-density-photon}, and the definitions $N_{\varphi(\gamma)} = \int_{V} d^{3}r~ n_{\varphi(\gamma)}$, where $V$ denotes the cloud volume, we obtain the evolution equations for the scalar and photon numbers in the superradiant cloud:
\begin{align}
\label{evolution-eq-l=1}
\begin{split}
    \frac{d N_{\varphi 00}}{dt} 
    &= 
    \Gamma_{\varphi} [-N_{\varphi 00} - AE_{00}^{1'}+ B_{1} F_{00}^{1'}] + \Gamma_{g}N_{\varphi 00}
    \,,  \\
    \frac{d N_{\gamma 00}}{dt} 
    &=
    \Gamma_{\varphi} [2N_{\varphi 00} +2AE_{00}^{1'}-2(B_{1}+B_{2})F_{00}^{1'} ]- \Gamma_{e} N_{\gamma 00}
    \,,  \\
    \frac{d N_{\varphi 20}}{dt} 
    &= 
    \Gamma_{\varphi} [-N_{\varphi 2 0} - AE_{20}^{1'}+ B_{1} F_{20}^{1'}] + \Gamma_{g}N_{\varphi 20}
    \,,  \\
    \frac{d N_{\gamma 20}}{dt} 
    &= 
    \Gamma_{\varphi} [2N_{\varphi 20} +2AE_{20}^{1'}-2(B_{1}+B_{2})F_{20}^{1'} ]- \Gamma_{e} N_{\gamma 20}
    \,,  \\
    \frac{d N_{\varphi 40}}{dt} 
    &= 
    \Gamma_{\varphi} [-N_{\varphi 40} - AE_{40}^{1'}+ B_{1} F_{40}^{1'}] 
    \,,  \\
    \frac{d N_{\gamma 40}}{dt} 
    &= 
    \Gamma_{\varphi} [2N_{\varphi 40} +2AE_{40}^{1'}-2(B_{1}+B_{2})F_{40}^{1'} ]- \Gamma_{e} N_{\gamma 40}
    \, ,
\end{split}
\end{align}
where $1'$ labels the $l=m=1$ mode, and $\Gamma_{g}$ is the growth rate of the scalar number density, determined via the continued fraction method as $\Gamma_{g} = 2\omega_{\text{I}}$.
In the above equations, we have defined
\begin{align}
    A=\frac{8\pi^{2}}{\beta_{\varphi} m_{\varphi}^{3}} \frac{1}{V} \,,\quad B_{1}=\frac{8\pi^{2}\beta_{\varphi}}{3 m_{\varphi}^{3}} \frac{1}{V} \,,\quad
    B_{2}=\frac{4\pi^{2}}{m_{\varphi}^{3}} \frac{1}{V} 
    \, ,
\end{align}
Using Eqs.~\eqref{eq-E}, \eqref{eq-F}, \eqref{eq-components}, and $N_{\varphi(\gamma)} = \int_{V} d^{3}r~ n_{\varphi(\gamma)}$, we obtain
\begin{align}
\begin{split}
    E_{00}^{1'} 
    &= 
    \frac{N_{\varphi 00}N_{\gamma 00}}{\sqrt{4\pi}} - \frac{N_{\varphi 00}N_{\gamma 20}}{\sqrt{20\pi}} 
    \,, \\
    E_{20}^{1'} 
    &=
    \frac{5 N_{\varphi 00}N_{\gamma 20}}{14 \sqrt{\pi}} - \frac{N_{\varphi 00}N_{\gamma 00}}{\sqrt{20\pi}} 
    \,, \\
    E_{40}^{1'} 
    &= 
    - N_{\varphi 00}N_{\gamma20} \sqrt{\frac{9}{245\pi}}
    \,, \\
    F_{00}^{1'} 
    &= 
    \frac{N_{\gamma 00}^{2} + N_{\gamma 20}^{2}}{\sqrt{4\pi}}
    \,, \\
    F_{20}^{1'} 
    &= 
    \frac{N_{\gamma 00} N_{\gamma 20}}{\sqrt{\pi}} + N_{\gamma 20}^{2}\sqrt{\frac{5}{49\pi}}
    \,, \\
    F_{40}^{1'} 
    &= 
    N_{\gamma 20}^{2}\sqrt{\frac{9}{49\pi}}
    \, .
\end{split}
\end{align}
The total scalar and photon numbers can thus be expressed as
\begin{align}
\begin{split}
    N_{\varphi (\gamma)} 
    &= 
    N_{\varphi (\gamma)00} Y_{00} +N_{\varphi (\gamma) 20} Y_{20} +N_{\varphi (\gamma)40} Y_{40}
    \\
    &= 
    N_{\varphi (\gamma)00} \frac{1}{\sqrt{4\pi}} + N_{\varphi (\gamma)20} \frac{1}{4}\sqrt{\frac{5}{\pi}} (3\cos^{2}\theta -1) 
    \\
    & \qquad
    +N_{\varphi (\gamma)40}\frac{3}{16}\sqrt{\frac{1}{\pi}} (35\cos^{4}\theta -30\cos^{2}\theta+3)
    \, .
\end{split}
\end{align}

\begin{figure}[H]
	\centering          
    \includegraphics[width=0.4\linewidth]{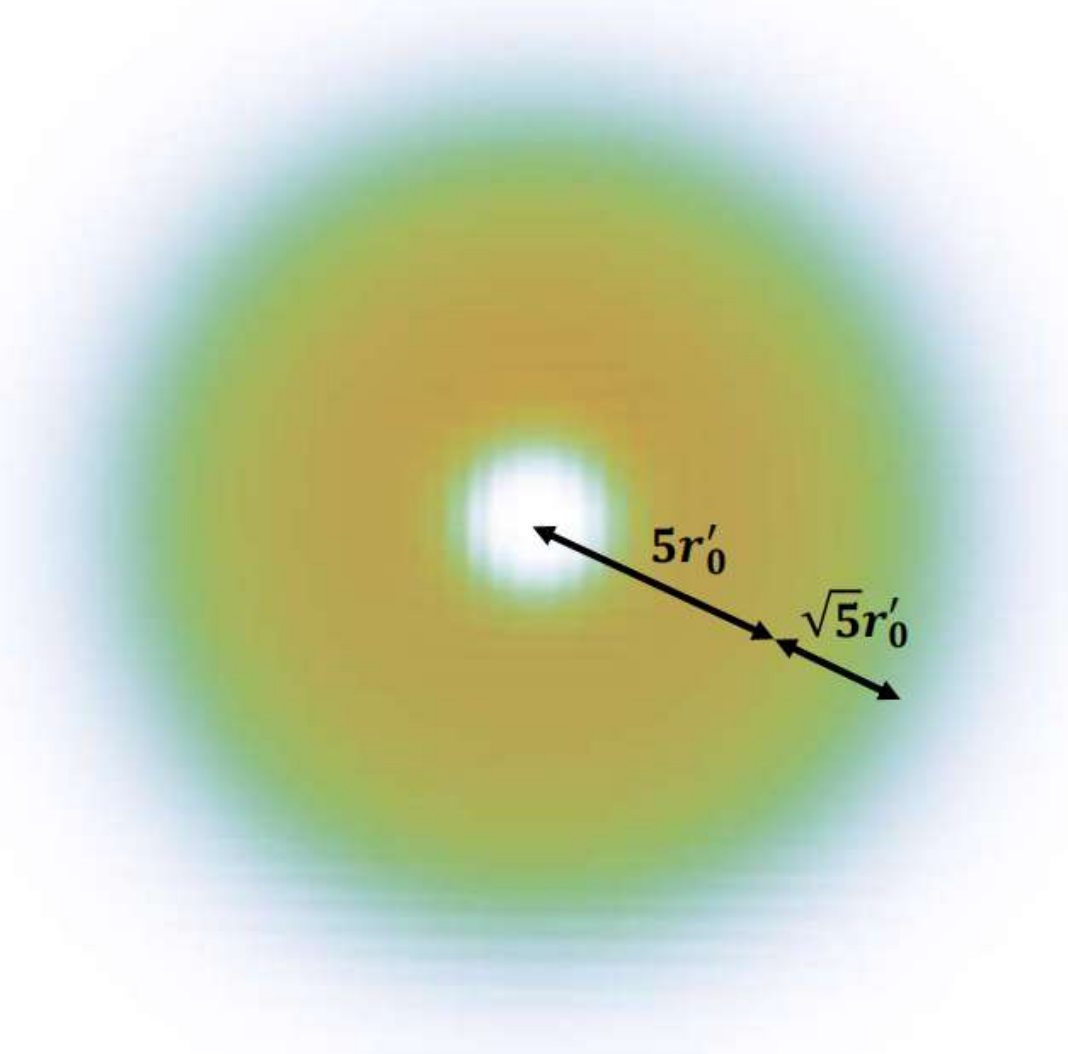}
	\caption{
    Model of the superradiant “2p” scalar cloud around a Kerr black hole.
    }
	\label{Fig1}
\end{figure}

For the ``2p'' model, as shown in Fig.~\ref{Fig1}, the corresponding superradiant cloud exhibits a toroidal structure, with mean radius $\langle r \rangle = 5r_{0}'$ and radial width $\Delta r = \sqrt{5}\, r_{0}'$, where $r_{0}'$ is defined as $r_0'=1/(\mu_0^2 M)$, and we have $V=2\pi^{2}\langle r \rangle \Delta r^{2}$, $\beta_{\varphi} = \mu_{0}M/2$, $\Gamma_{e} = 1/\Delta r$. 
We adopt $a=0.7M$, $\mu_{0}M=0.1$, and $\mu_{0}=10^{-5}\,~\mathrm{eV}$, which yield
\begin{align}\label{uniform-1}
\begin{split}
    M
    &= 
    2.66 \times 10^{24}~\mathrm{kg} 
    \,,\quad 
    \Gamma_{\varphi} 
    = 
    (3\times10^{40})^{-1}~\mathrm{s}^{-1}
    \,, \\
    \Gamma_{g} 
    &= 
    1.67~\mathrm{s}^{-1}
    \,,\quad
    \Gamma_{e} 
    = 
    6.8 \times 10^{8}~\mathrm{s}^{-1}
    \, .
\end{split}
\end{align}
Note that the above value of $\Gamma_{\varphi}$ follows from our choice $\mu_{0}=10^{-5}~\mathrm{eV}$ and $\beta_{\gamma}\simeq 10^{7}$.
Throughout this work, we fix $\beta_{\gamma}=10^{7}$ as a benchmark value.
In our analysis of the stimulated decay of the superradiant scalar cloud, the black hole masses considered are smaller than the Earth's mass.
Such a small mass scale is naturally associated with primordial black holes (PBHs)~\cite{Carr:2009jm, Carr:2017jsz, Qiu:2022klm}.
PBHs can merge to form more massive black holes with nonzero spin, which can subsequently host the formation of superradiant clouds~\cite{Scheel:2008rj}.
The numerical solutions of Eq.~\eqref{evolution-eq-l=1} for this parameter set are summarized below.

\begin{figure}[H]
	\centering
	\includegraphics[width=0.45\linewidth]{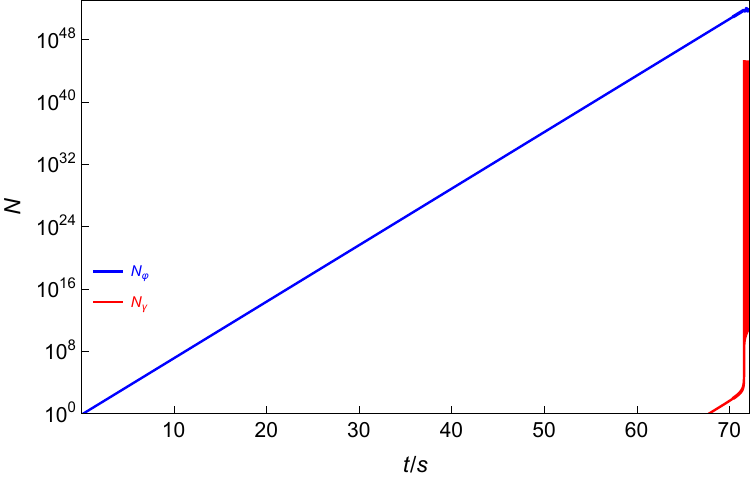}
    \includegraphics[width=0.45\linewidth]{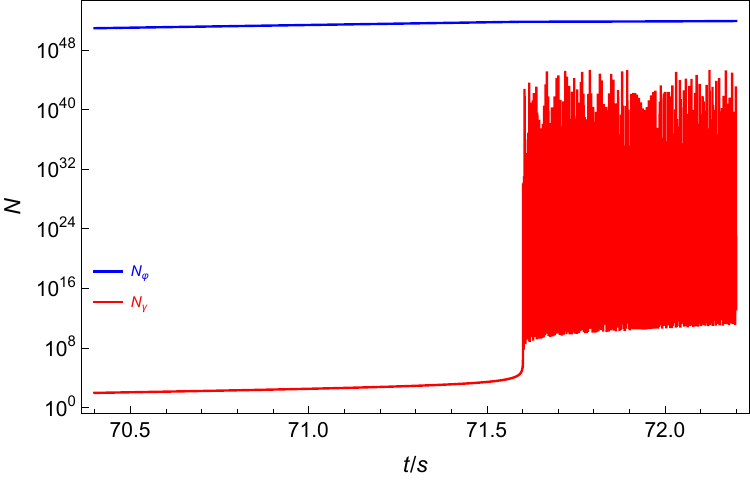}
	\caption{
    The evolution of scalar (blue line) and photon (red line) number in the non-spherical for $l=m=1$. 
    The figure is generated based on the parameter values given in Eq.~\eqref{uniform-1}.
    The left panel shows the entire time scale considered here, while the right panel presents an enlarged view around the time at which the photon number begins to increase.
    } 
	\label{l=1-N}
\end{figure}

\begin{figure}[H]
	\centering
	\subfigure{
		\includegraphics[width=0.45\linewidth]{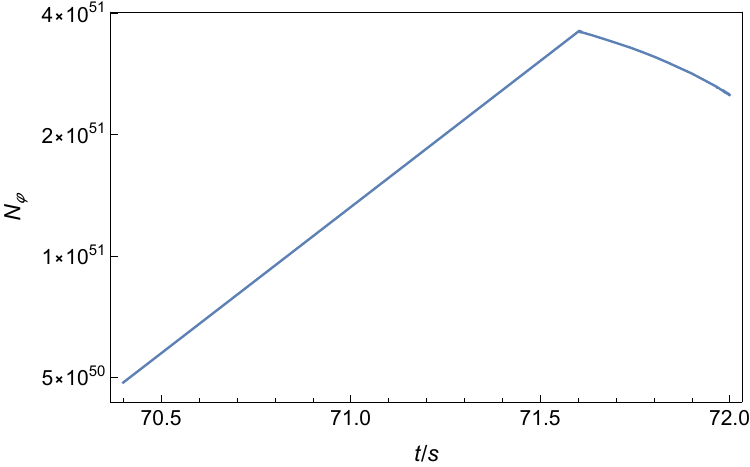}
	}
	\subfigure{
		\includegraphics[width=0.45\linewidth]{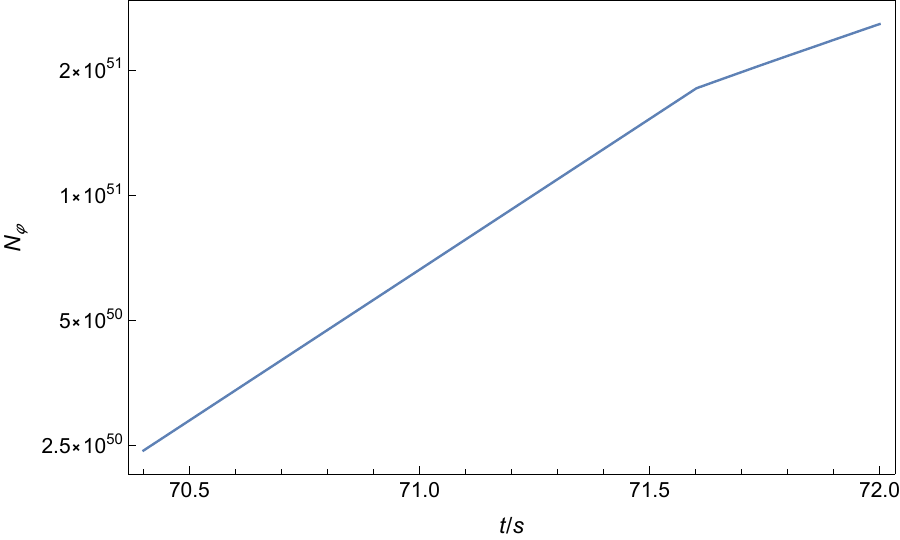}
	}
	\subfigure{
		\includegraphics[width=0.45\linewidth]{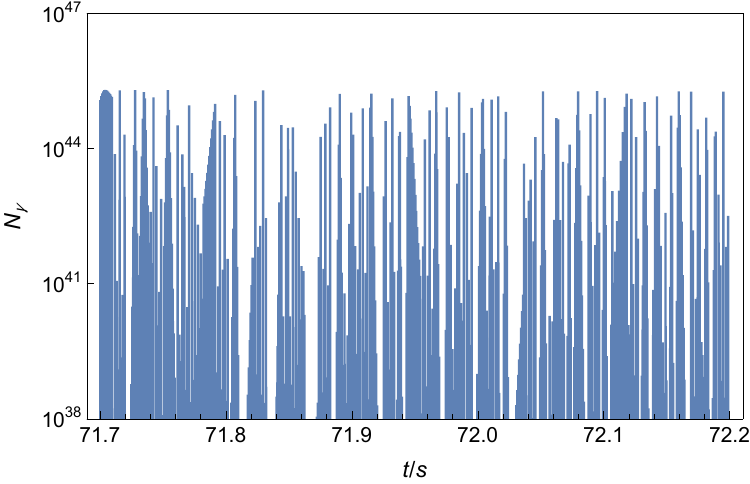}
        }
        \subfigure{
		\includegraphics[width=0.45\linewidth]{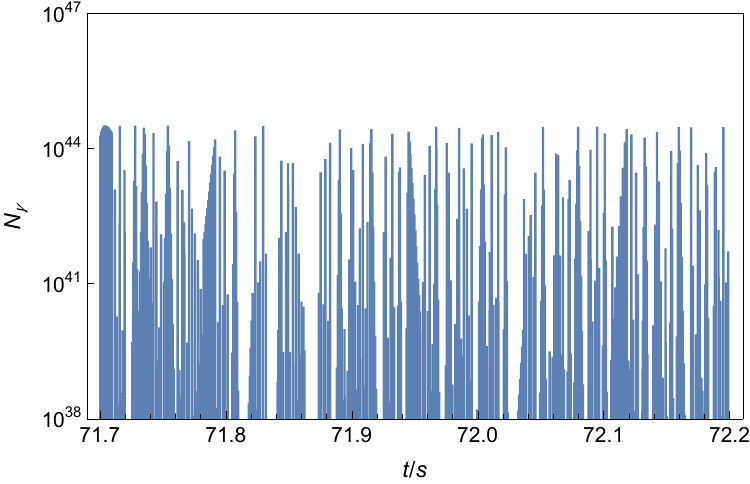}
	}
    \caption{
    scalar (top panels) and photon (bottom panels) number in the non-spherical for different angles (left panel:$\theta = \pi/2$, right panel:$\theta = \pi/4$).
    }
	\label{l=1-N-pi}
\end{figure}

\begin{figure}[H]
	\centering
	\includegraphics[width=0.45\linewidth]{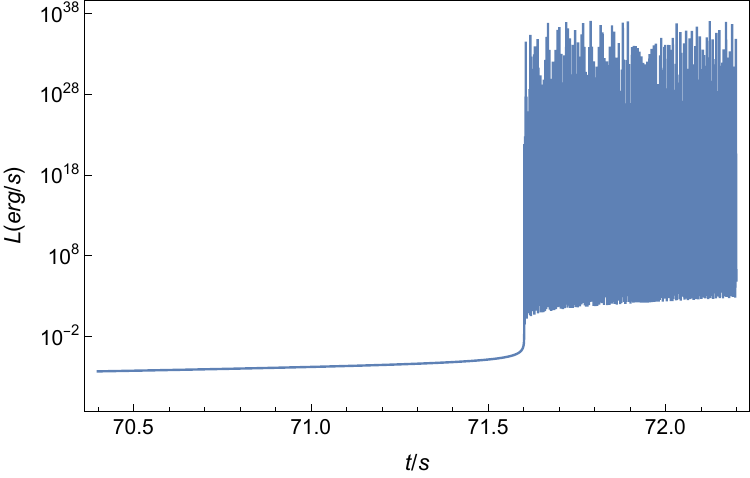}
    \includegraphics[width=0.45\linewidth]{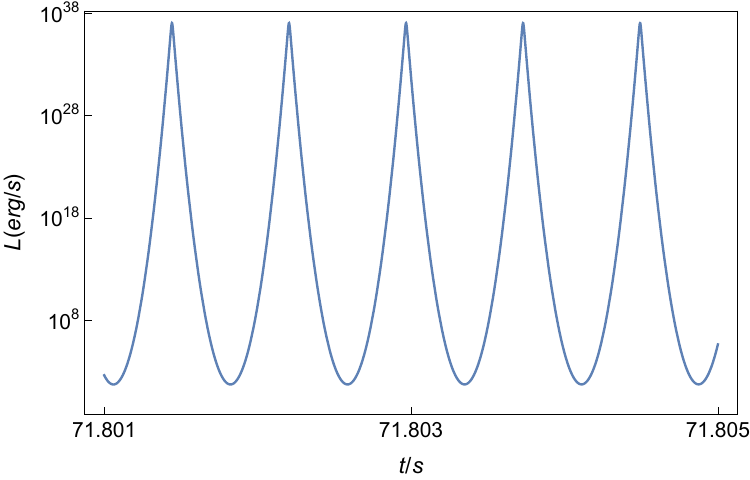}
	\caption{
    The evolution of the average photon luminosity after the stimulated decay takes over.
    }
    \label{l=1-L}
\end{figure}

\noindent 
In Fig.~\ref{l=1-N}, due to the superradiant instability, the scalar number initially grows exponentially until stimulated decay becomes dominant.
At this stage, the photon number rises sharply and subsequently exhibits oscillatory behavior.
These oscillations arise from the alternating dominance between stimulated scalar decay and photon escape from the cloud.
\noindent
Figure~\ref{l=1-N-pi} shows the angular dependence of the scalar and photon numbers.
Both quantities peak near the equatorial plane, where the emission is strongest, and gradually decrease toward the polar regions.
At the equatorial plane $(\theta=\pi/2)$, once stimulated emission sets in, the scalar decay rate exceeds the replenishment rate from superradiant growth, leading to a net reduction in the scalar number.
In contrast, away from the equatorial plane, stimulated emission merely suppresses the growth rate rather than reversing it.
As a result, the scalar and photon distributions tend toward isotropization as the system evolves. 
Figure~\ref{l=1-L} presents the luminosity of the emitted radiation, defined as $L=(\mu_0/2)N_{\gamma}\Gamma_{e}$.
We find that stimulated emission from the superradiant scalar cloud can reach luminosities of order $\sim 10^{37}\,\mathrm{erg/s}$.
Such luminosities are comparable to those of compact astrophysical transients, indicating that superradiant scalar clouds may act as sources of intense and coherent radiation in astrophysical environments.

\subsection{$l=2, m=1$}

Although the superradiant growth rate of the $(l=2,m=1)$ mode is smaller than that of the $(l=m=1)$ mode, a comparative analysis of the stimulated radiation produced by different superradiant scalar clouds is still instructive.
In particular, differences in angular structure and mode-dependent cloud profiles can lead to distinct observational signatures.

For $(l=2,m=1)$, the scalar number density is proportional to
$Y_{2}^{\pm 1*} Y_{2}^{\pm 1} \propto \cos^{2} \theta \sin^{2}\theta$,
which yields
\begin{align}
\begin{split}
    n_{\varphi} 
    &= 
    \Theta(R-r) n_{\varphi}(t) \cos^{2}\theta\sin^{2}\theta
    \, ,
    \\
    &= 
    \Theta(R-r) n_{\varphi}(t) 
    \cdot 
    \frac{4\sqrt{\pi}}{3} 
    \left(Y_{00}-\frac{1}{\sqrt{5}}Y_{20}\right)
    \cdot
    \frac{4\sqrt{\pi}}{3} 
    \left(Y_{00}+\frac{2}{\sqrt{5}}Y_{20}\right)
    \\
    &= 
    \Theta(R-r) [n_{\varphi 00}(t) Y_{00} + n_{\varphi 20} Y_{20} + n_{\varphi 40} Y_{40}]
    \, ,
\end{split}
\end{align}
from which one obtains $n_{\varphi 20}(t) = \sqrt{5}/7\, n_{\varphi 00}(t)$ and $n_{\varphi 40}(t) = - 4/7\, n_{\varphi 00}(t)$.
To avoid unphysical behavior in the truncated angular expansion, an additional $n_{\varphi 60}$ and $n_{\gamma 60}$ correction term is included.
Thus, the scalar and photon number densities can be expanded as
\begin{align}
\begin{split}
    n_{\varphi} 
    &=
    n_{\varphi 00}Y_{00} +n_{\varphi 20}Y_{20} + n_{\varphi 40}Y_{40} + n_{\varphi 60}Y_{60}
    \,, \\
    n_{\gamma} 
    &=
    n_{\gamma 00}Y_{00} +n_{\gamma 20}Y_{20} + n_{\gamma 40}Y_{40} 
    + n_{\gamma 60}Y_{60}
    \, .
\end{split}
\end{align}
The corresponding evolution equations for the scalar and photon numbers are given by
\begin{align}
\label{evlutoin-l=2}
\begin{split}
    \frac{d N_{\varphi 00}}{dt} 
    &= 
    \Gamma_{\varphi} [-N_{\varphi 00} - AE_{00}^{2'}+ B_{1} F_{00}^{2'}] + \Gamma_{g}N_{\varphi 00}
    \, , \\
    \frac{d N_{\gamma 00}}{dt} 
    &=
    \Gamma_{\varphi} [2N_{\varphi 00} +2AE_{00}^{2'}-2(B_{1}+B_{2})F_{00}^{2'} ]- \Gamma_{e} N_{\gamma 00}
    \, , \\
    \frac{d N_{\varphi 20}}{dt} 
    &= 
    \Gamma_{\varphi} [-N_{\varphi 2 0} - AE_{20}^{2'}+ B_{1} F_{20}^{2'}] + \Gamma_{g}N_{\varphi 20}
    \, , \\
    \frac{d N_{\gamma 20}}{dt} 
    &=
    \Gamma_{\varphi} [2N_{\varphi 20} +2AE_{20}^{2'}-2(B_{1}+B_{2})F_{20}^{2'} ]- \Gamma_{e} N_{\gamma 20}
    \, , \\
    \frac{d N_{\varphi 40}}{dt} 
    &= 
    \Gamma_{\varphi} [-N_{\varphi 40} - AE_{40}^{2'}+ B_{1} F_{40}^{2'}] + \Gamma_{g}N_{\varphi 20}
    \, , \\
    \frac{d N_{\gamma 40}}{dt} 
    &=
    \Gamma_{\varphi} [2N_{\varphi 40} +2AE_{40}^{2'}-2(B_{1}+B_{2})F_{40}^{2'} ]- \Gamma_{e} N_{\gamma 40}
    \, , \\
    \frac{d N_{\varphi 60}}{dt} 
    &= 
    \Gamma_{\varphi} [-N_{\varphi 60} - AE_{60}^{2'}+ B_{1} F_{60}^{2'}]
    \, , \\
    \frac{d N_{\gamma 60}}{dt} 
    &=
    \Gamma_{\varphi} [2N_{\varphi 60} +2AE_{60}^{2'}-2(B_{1}+B_{2})F_{60}^{2'} ]- \Gamma_{e} N_{\gamma 60}
    \, .
\end{split}
\end{align}
Where $2'$ labels the $l=2,m=1$ mode, the explicit forms of $E_{l0}^{2'}$ and $F_{l0}^{2'}$ 
are given by
\begin{align}
\begin{split}
    E_{00}^{2'} 
    &= 
    \frac{N_{\varphi 00}N_{\gamma 00}}{\sqrt{4\pi}}
    + \frac{\sqrt{5}N_{\varphi 00}N_{\gamma 20}}{14\sqrt{\pi}} 
    - \frac{2N_{\varphi 00}N_{\gamma 40}}{7\sqrt{\pi}}
    \, , \\
    E_{20}^{2'} 
    &= 
    \frac{\sqrt{5} N_{\varphi 00}N_{\gamma 00}}{14 \sqrt{\pi}}
    + \frac{5 N_{\varphi 00}N_{\gamma 20}}{14\sqrt{\pi}} 
    - \frac{\sqrt{5} N_{\varphi 00}N_{\gamma 40}}{77\sqrt{\pi}}
    \, , \\
    E_{40}^{2'} 
    &= 
    -\frac{2 N_{\varphi 00}N_{\gamma 00}}{7 \sqrt{\pi}}
    - \frac{\sqrt{5}N_{\varphi 00}N_{\gamma 20}}{77 \sqrt{\pi}} 
    + \frac{909 N_{\varphi 00}N_{\gamma 40}}{2002\sqrt{\pi}}
    \, , \\
    E_{60}^{2'} 
    &= 
    - \frac{30 \sqrt{5}N_{\varphi 00}N_{\gamma 20}}{77\sqrt{13\pi}}
    - \frac{5 N_{\varphi 00} N_{\gamma 40}}{154 \sqrt{13\pi}}
    \, , \\
    F_{00}^{2'} 
    &= 
    \frac{N_{\gamma 00}^{2}+N_{\gamma 20}^{2}+N_{\gamma 40}^{2}}{\sqrt{4\pi}}
    \, , \\
    F_{20}^{2'} 
    &=
    \frac{N_{\gamma 00} N_{\gamma 20}}{\sqrt{\pi}}
    + \frac{\sqrt{5}N_{\gamma 20}^{2}}{7\sqrt{\pi}} 
    + \frac{6N_{\gamma 00}N_{\gamma 20}}{7\sqrt{\pi}}
    + \frac{10\sqrt{5}N_{\gamma 40}^{2}}{77\sqrt{\pi}}
    \, , \\
    F_{40}^{2'} 
    &= 
    \frac{N_{\gamma 00}N_{\gamma 40}}{\sqrt{\pi}}
    +\frac{3N_{\gamma 20}^{2}}{7\sqrt{\pi}}
    + \frac{20\sqrt{5}N_{\gamma 20}N_{\gamma 40}}{77\sqrt{\pi}}
    + \frac{243 N_{\gamma 40}^{2}}{1001 \sqrt{\pi}}
    \, , \\
    F_{60}^{2'} 
    &=
    \frac{15\sqrt{5}N_{\gamma 20}N_{\gamma 40}}{11\sqrt{13\pi}}
    + \frac{10 N_{\gamma 40}^{2}}{11 \sqrt{13\pi}}
   \, ,
\end{split}
\end{align}
while the coefficients $A$, $B_{1}$, and $B_{2}$ are defined in the same way as in the $(l = m = 1)$ model.

\begin{figure}[H]
	\centering          
    \includegraphics[width=0.45\linewidth]{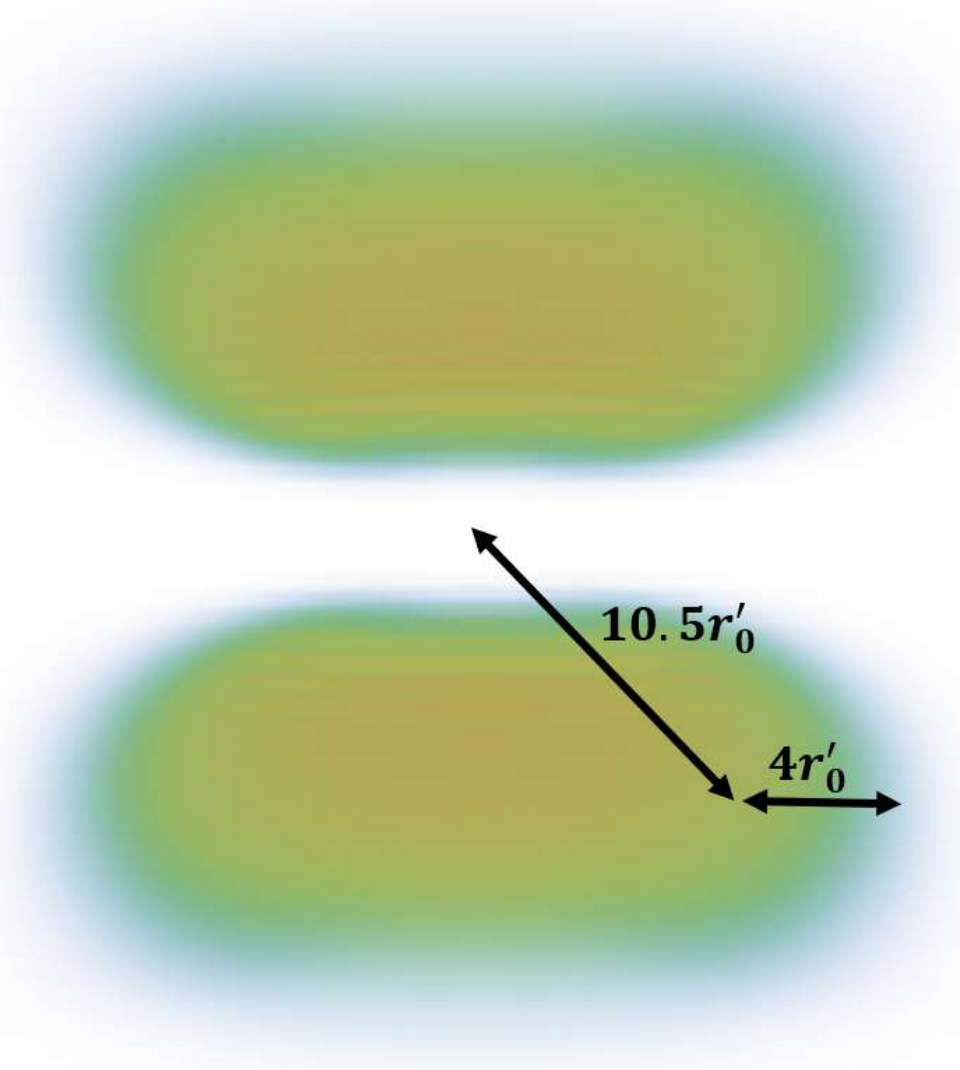}
	\caption{
    Model of the superradiant “$l=2,m=1$” scalar cloud around a Kerr black hole.
    }
	\label{l=2}
\end{figure}

The spatial structure of the $(l=2,m=1)$ superradiant scalar cloud is illustrated in Fig.~\ref{l=2}.
We adopt $\langle r \rangle = 10.5r_{0}'$ and $\Delta r = 4 r_{0}'$, with
$V \approx 4\pi/3 (r_{2}^{3}-r_{1}^{3})$,
where $r_{1} = \langle r \rangle - \Delta r$ and $r_{2} = \langle r \rangle + \Delta r$.
We further take $\beta_{\varphi} = \mu_{0}M/3$ and $\Gamma_{e} = 1/\Delta r$.
For the numerical analysis, we choose $a=0.7 M$, $\mu_{0} M= 0.1$, and $\mu_{0}=10^{-5} ~\mathrm{eV}$, which correspond to
\begin{align}\label{uniform-2}
\begin{split}
    M 
    &= 2.66 \times 10^{24}~\mathrm{kg} 
    \,,\quad 
    \Gamma_{\varphi} 
    = 
    (3\times10^{40})^{-1}~\mathrm{s}^{-1}
    \,, \\
    \Gamma_{g} 
    &= 
    2\times10^{-3}~\mathrm{s}^{-1}
    \,,\quad
    \Gamma_{e} = 3.8 \times 10^{8}~\mathrm{s}^{-1}
    \, .
\end{split}
\end{align}

\begin{figure}[H]
	\centering
	\includegraphics[width=0.4\linewidth]{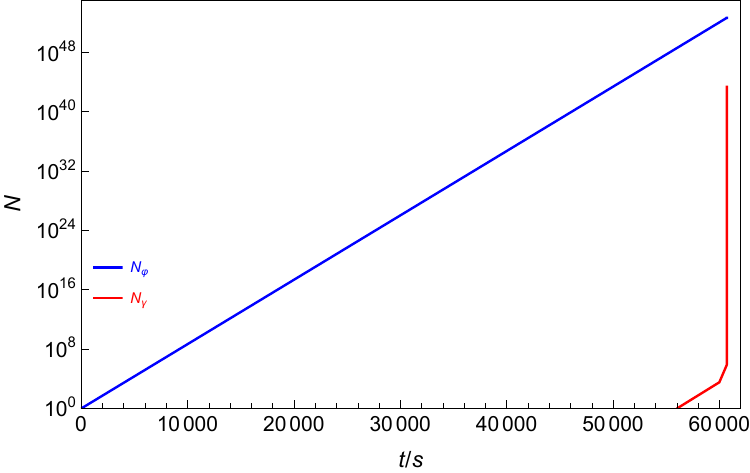}
    \includegraphics[width=0.4\linewidth]{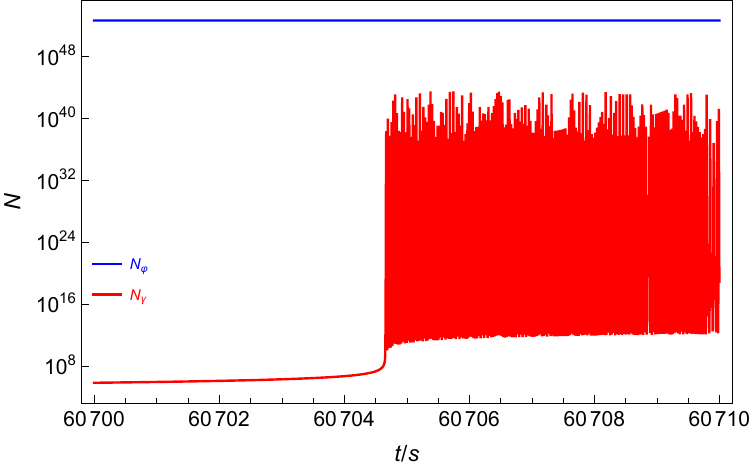}
	\caption{
    The evolution of scalar (blue line) and photon (red line) number in the non-spherical for $l=2,m=1$.
    The figure is generated based on the parameter values given in Eq.~\eqref{uniform-2}.
    }
    \label{l=2-N}
\end{figure}

\begin{figure}[H]
	\centering
	\subfigure{
		\includegraphics[width=0.4\linewidth]{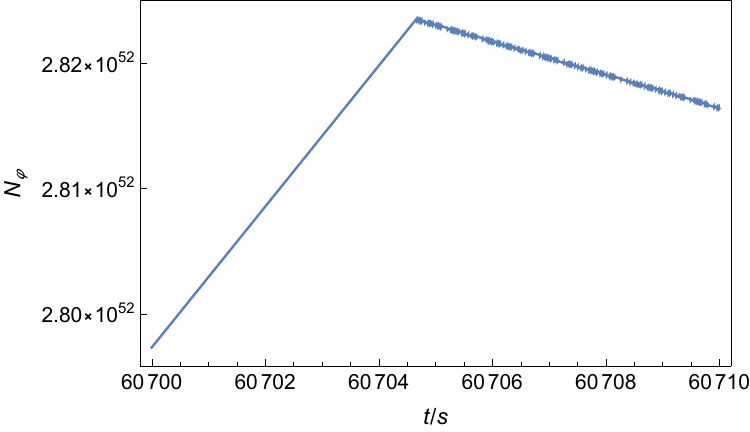}
	}
	\subfigure{
		\includegraphics[width=0.4\linewidth]{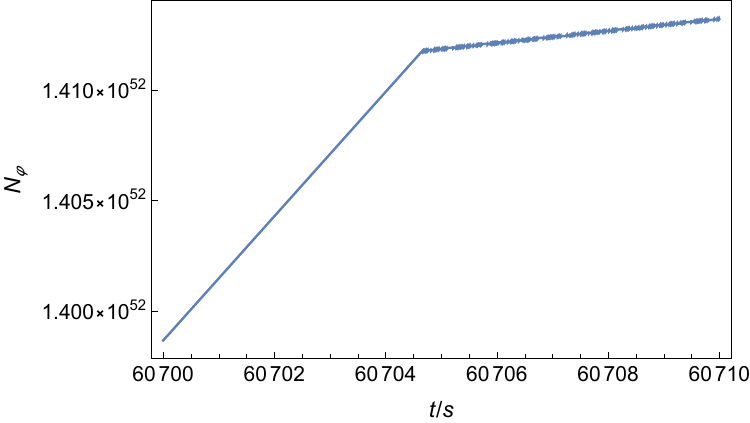}
	}
	\subfigure{
		\includegraphics[width=0.4\linewidth]{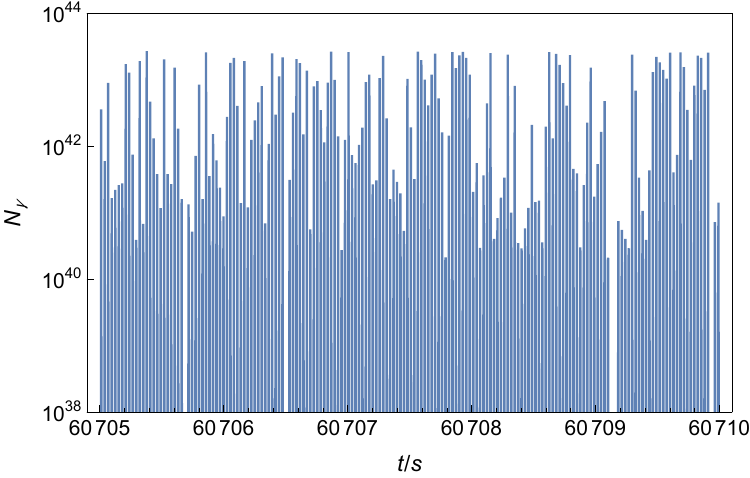}
        }
        \subfigure{
		\includegraphics[width=0.4\linewidth]{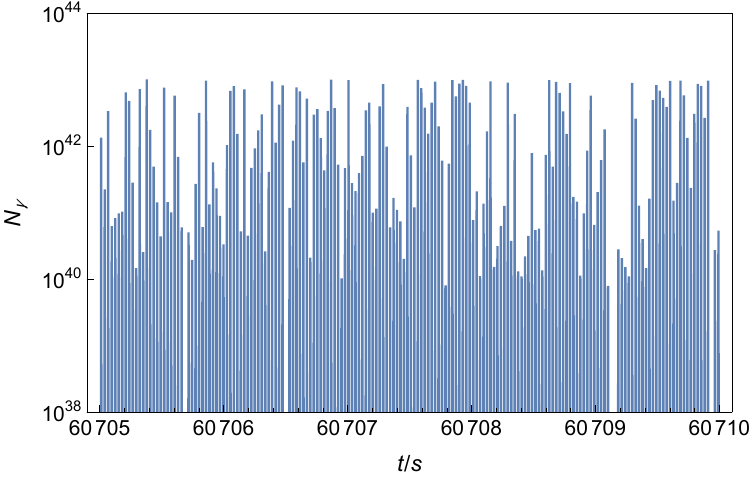}
	}	
    \caption{
    The scalar (top panels) and photon (bottom panels) number in the non-spherical for different angles (left panel:$\theta = \pi/4$, right panel:$\theta = \pi/8$).
    }
	\label{l=2-N-pi}
\end{figure}

\begin{figure}[H]
	\centering
	\includegraphics[width=0.45\linewidth]{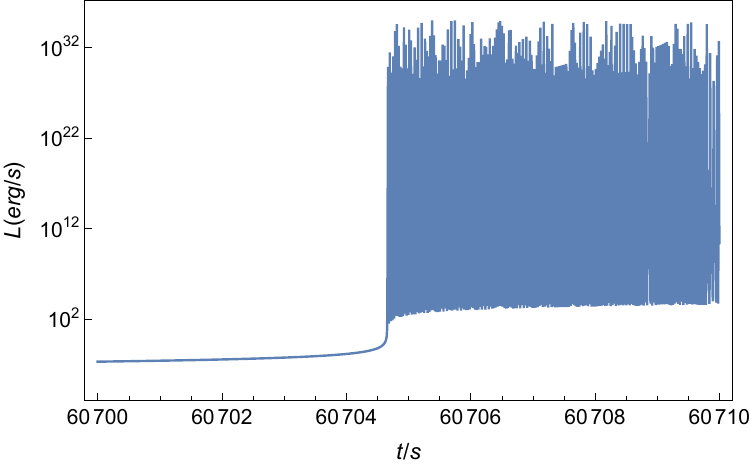}
    \includegraphics[width=0.45\linewidth]{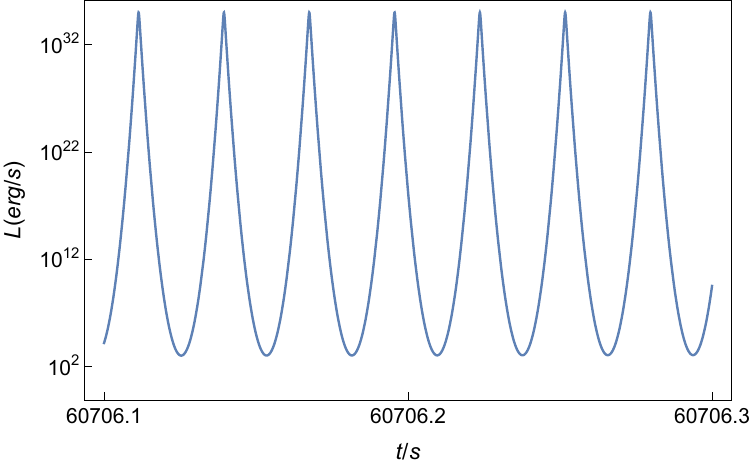}
	\caption{
    The evolution of the average photon luminosity after the stimulated decay takes over.
    }
    \label{l=2-L}
\end{figure}

\begin{figure}[H]
	\centering
	\includegraphics[width=0.45\linewidth]{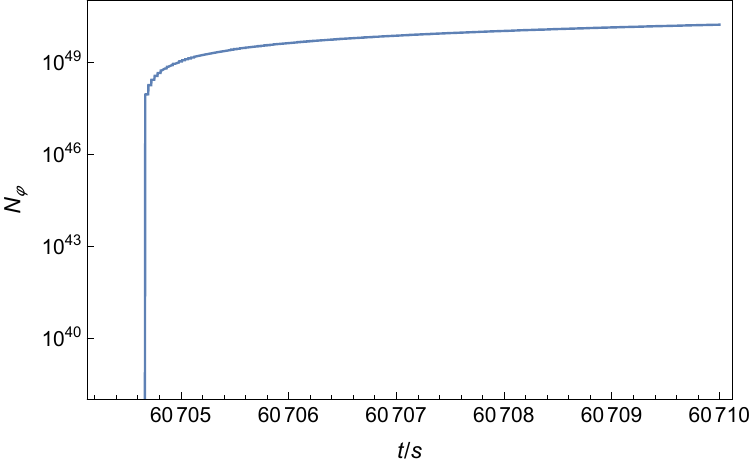}
    \includegraphics[width=0.45\linewidth]{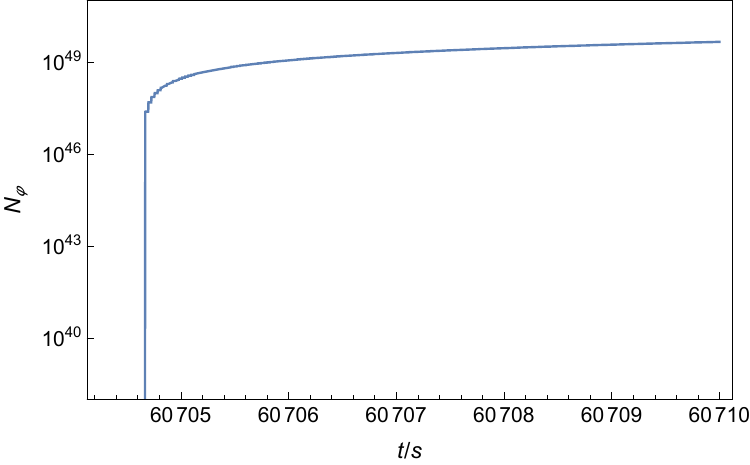}
	\caption{
    The evolution of the scalar number for $\theta = 0$ (left) and $\theta = \pi/2$ (right).
    }
    \label{l=2-Ns-0}
\end{figure}

\noindent 
The numerical solutions of Eq.~\eqref{evlutoin-l=2} are shown in
Figs.~\ref{l=2-N}–\ref{l=2-N-pi}.
Overall, the time evolution of the scalar and photon numbers exhibits qualitative similarities to the $(l=m=1)$ case.
However, owing to the different angular structure of the $(l=2,m=1)$ mode, the stimulated emission is maximal at $\theta = \pi/4$. Moreover, the reduced superradiant growth rate and the broader spatial profile of the cloud lead to noticeable differences in both the intensity and duration of the stimulated radiation. 
\noindent
As shown in Fig.~\ref{l=2-L}, the photon luminosity $L=(\mu_0/2) N_\gamma \Gamma_e$ reaches $\sim 10^{35}\,\mathrm{erg/s}$ after the stimulated decay dominates, with a longer emission timescale than in the $(l=m=1)$ case.
During the stimulated emission phase, the scalar number increases at both $\theta = 0$ and $\theta = \pi/2$, as illustrated in Fig.~\ref{l=2-Ns-0}.
This behavior indicates that the stimulated process partially erases the initial angular structure of the cloud, leading to an effective angular redistribution of the scalar field.
In the case of a uniform matter distribution, the superradiant instability of the scalar field closely resembles that of other bosonic fields, where the mass is a constant for both cases.
If other bosons (such as axions or axion-like particles) have the same spontaneous decay rate at the same mass, their stimulated emission exhibits similar behavior.
Consequently, stimulated radiation alone cannot be used to distinguish between these scenarios.
In Sec.~\ref{non-uniform}, we extend the analysis to non-uniform matter distributions, for which the effective mass of the scalar field is no longer a constant due to the chameleon mechanism. This essential difference from the uniform matter distribution case leads to a qualitatively distinct and enhanced behavior of stimulated emission from superradiant scalar clouds.

\section{The non-uniform distribution of matter}
\label{non-uniform}

To investigate the effects of a matter distribution, namely the chameleon mechanism, we consider the non-uniform matter distribution.
We consider a case where $\mu_{0}^{2} \neq 0$, $F(\theta)=0$, and $G(r)= \beta \Theta[r-r_{0}] (r-r_{0}) r^{-n}$ as a benchmark model.
Here, $r_{0}$ denotes the distance from the center of the black hole to the inner edge of the matter distribution.
A parameter $n$ characterizes the radial decay rate of the matter, and $\beta$ represents the coupling strength between matter and the scalar field.
This setup provides a toy model of a geometrically thick accretion disk~\cite{Cardoso:2013fwa}.
Assuming the matter can be described as a pressureless perfect fluid, the trace of the stress-energy tensor is approximated by $T_{1}(r,\theta) \approx - \rho(r,\theta)$, where the matter density $\rho(r,\theta)$ takes the form
\begin{equation}
    \rho(r,\theta)=\dfrac{A_0}{2A_2}\beta \cdot \dfrac{\Theta[r-r_{0}] (r-r_{0}) r^{-n}}{r^{2} + a^{2} \cos^{2}\theta}.
\end{equation}
Throughout this section, we restrict our analysis to the $l=m=1$ mode.

\subsection{Superradiant instability}

We determine the bound-state eigenfrequencies by solving the radial equation for $R_{lm}(r)$ using the direct integration (shooting) method~\cite{Berti:2009kk, Molina:2010fb, Zhang:2014kna}.
The physical boundary conditions for the radial function are given by
\begin{align}\label{eq:R_boundary}
    R_{lm}(r) \;\longrightarrow\;
    \begin{cases} 
    (r-r_+)^{- i \sigma}
    & (r \to r_+)
    \, ,
    \\
    e^{qr}~r^{( \mu_{0}^2-2\omega^{2})M/q - 1}
    & (r \to \infty) 
    \, ,
 \end{cases}
\end{align}
where $\sigma=2 r_+ M (\omega - \omega_\text{c})/(r_+ - r_-)$, $q=-\sqrt{\mu_{0}^{2}-\omega^{2}}$ and $\omega_\text{c} \equiv m \Omega_\text{H}$.
Since both the event horizon and spatial infinity correspond to singular points of the radial equation, the asymptotic behaviors in Eq.~\eqref{eq:R_boundary} cannot be imposed directly as numerical initial conditions. 
Instead, we perform series expansions at both boundaries to construct well-defined solutions suitable for numerical integration.

Near the event horizon $r=r_+$, the radial equation admits a regular singular point. 
Imposing the purely ingoing boundary condition, we express the radial function in a Frobenius-expanded form,
\begin{align}
\label{eq:horizon_series}
    R_{lm}(r)
    =
    (r-r_+)^{- i \sigma}
    \sum_{n=0}^{\infty} a_n (r-r_+)^n 
    \, ,
\end{align}
where the coefficients $a_n$ are determined recursively by substituting this expansion into the radial equation. 
At spatial infinity, the radial function is expanded as
\begin{align}
\label{eq:infinity_series}
    R_{lm}(r)
    =
    e^{qr}~r^{( \mu_{0}^2-2\omega^{2})M/q - 1}
    \sum_{n=0}^{\infty} \frac{b_n}{r^n}
    \, ,
\end{align}
with the coefficients $b_n$ again fixed order by order.

\begin{figure}[H]
	\centering		
    \includegraphics[width=0.45\linewidth]{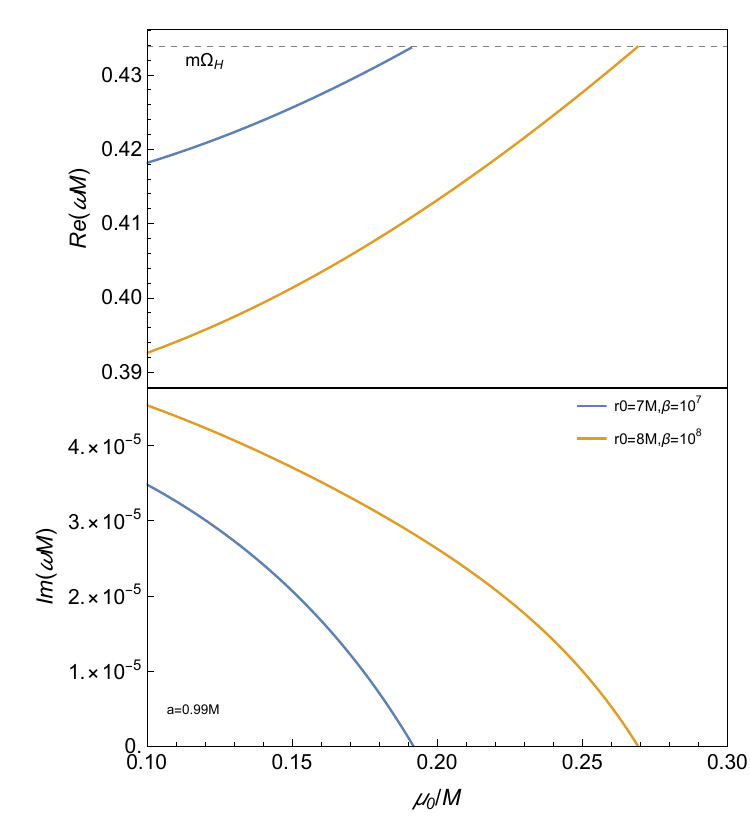}
    \includegraphics[width=0.45\linewidth]{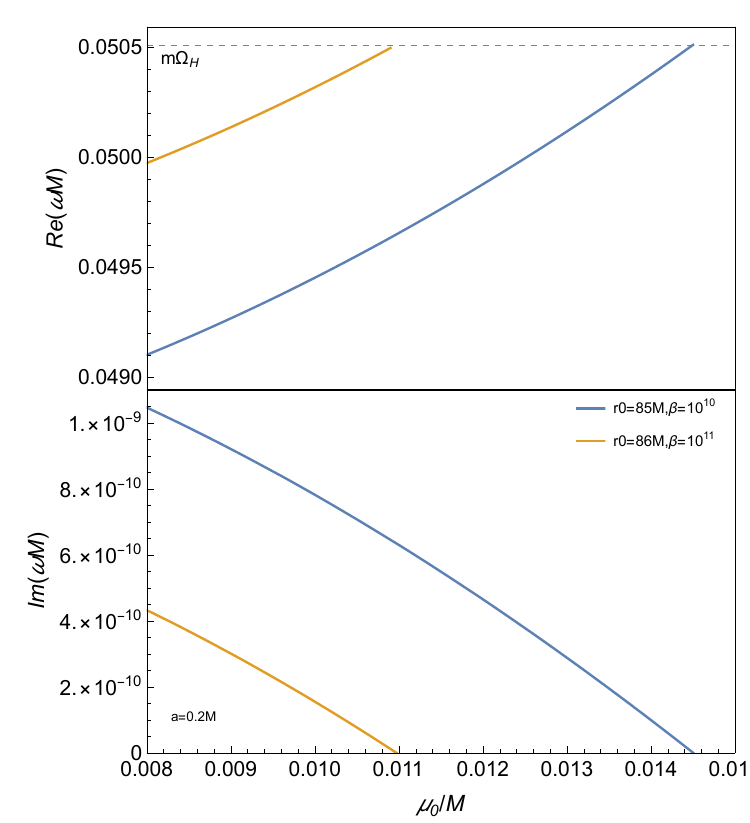}
	\caption{
    Superradiant instability of a scalar field as a function of $\mu_{0}$ in a non-uniform matter background, with $n=4$. Left panel: $a=0.99M$; right panel: $a=0.2M$.
    }
    \label{mu}
\end{figure}

With the series-expanded boundary conditions specified above, we determine the eigenfrequencies using a shooting-and-matching procedure. 
Specifically, we first integrate the radial equation outward from the black hole event horizon to an intermediate matching point at $r = r_\text{m}$. 
Independently, we integrate the radial equation inward from spatial infinity to the same matching point $r = r_\text{m}$. 
The eigenfrequencies of the perturbations are then obtained by requiring that the two solutions match smoothly at $r=r_\text{m}$.
The resulting numerical spectra are summarized below.
As shown in Fig.~\ref{mu}, the real part of the frequency $\omega_{\text{R}}$ increases with the scalar field mass $\mu_{0}$, which is consistent with the behavior of superradiant instabilities for other massive bosonic fields. 
In contrast to the uniform matter case, however, the efficiency of the scalar superradiant instability can be significantly enhanced in the presence of non-uniform matter distributions.

\begin{figure}[H]
	\centering		
    \includegraphics[width=0.45\linewidth]{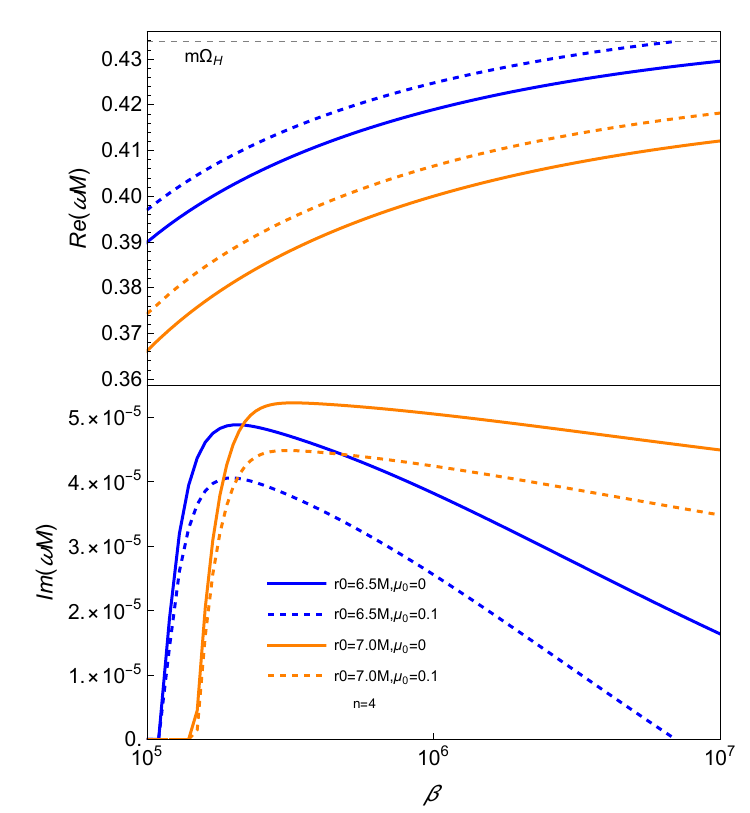}        
    \includegraphics[width=0.45\linewidth]{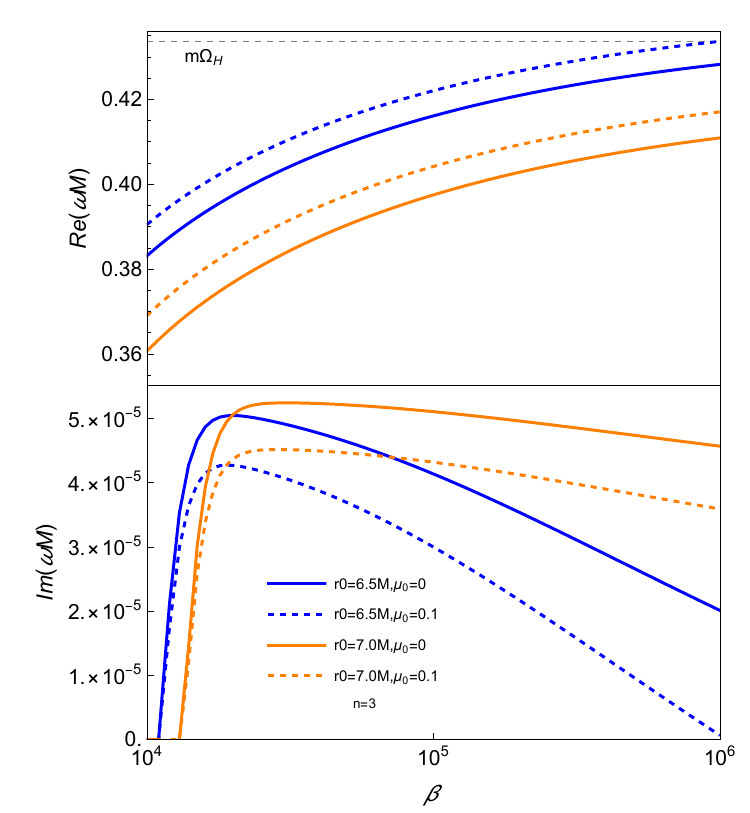}
	\caption{
    Superradiant instability of a scalar field as a function of $\beta$ in a non-uniform matter background, with $a=0.99M$.
    Left panel: $n=4$; right panel: $n=3$.
    }
    \label{n}
\end{figure}

\begin{figure}[H]
	\centering		
    \includegraphics[width=0.45\linewidth]{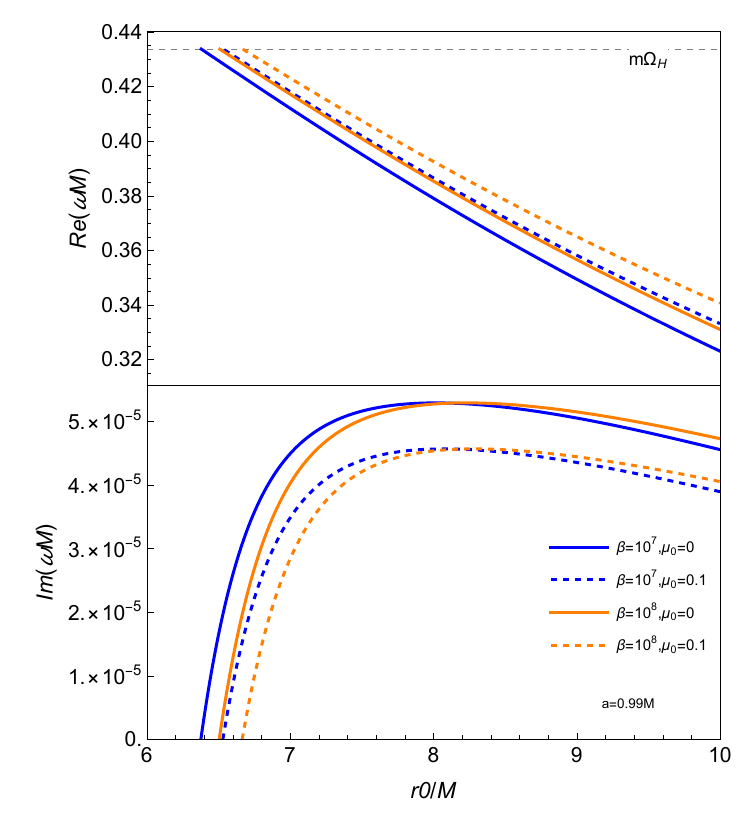}        
    \includegraphics[width=0.45\linewidth]{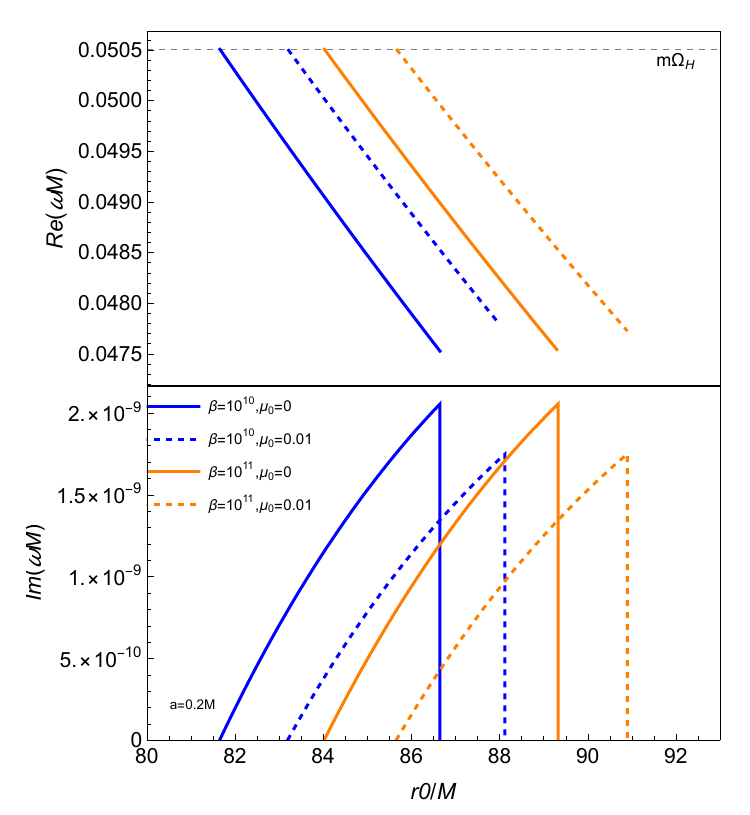}
	\caption{
    Superradiant instability of a scalar field as a function of $r_{0}$ in a non-uniform matter background, with $n=4$.
    Left panel: $a=0.99M$; right panel: $a=0.2M$.
    }
    \label{r0}
\end{figure}

\begin{figure}[H]
	\centering		
    \includegraphics[width=0.45\linewidth]{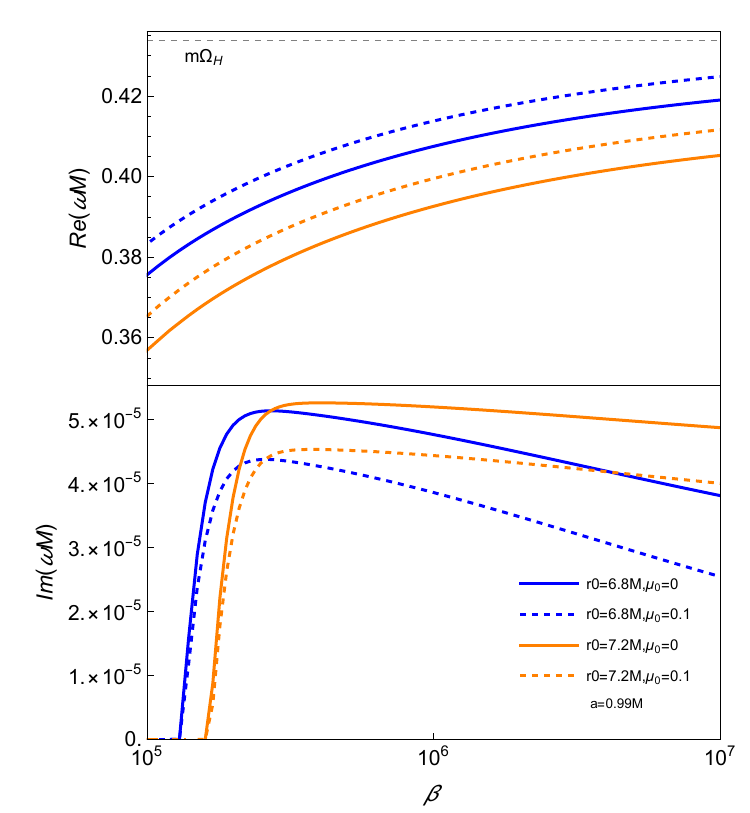}        
    \includegraphics[width=0.45\linewidth]{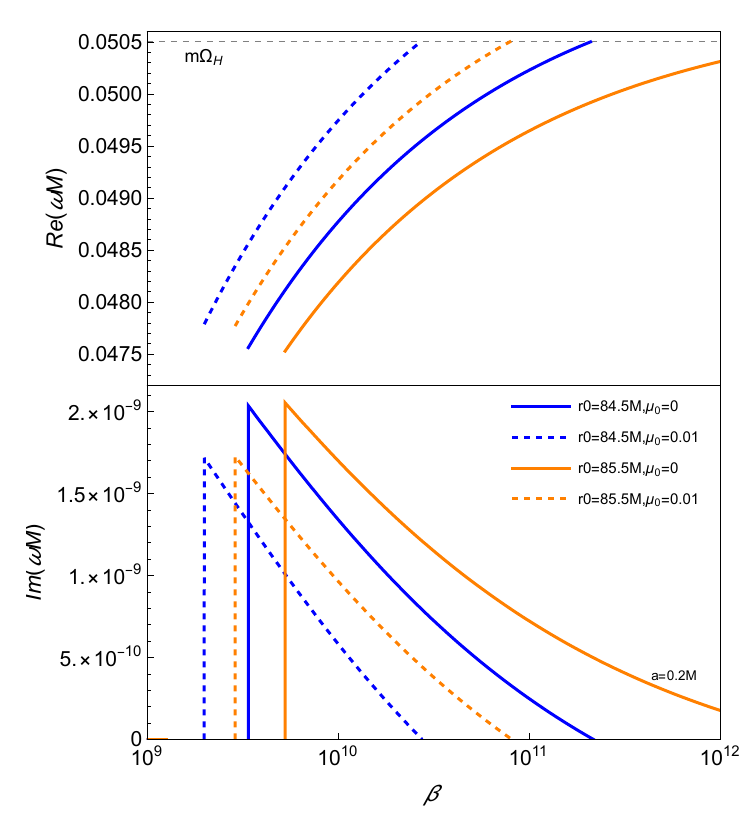}
	\caption{
    Superradiant instability of a scalar field as a function of $\beta$ in a non-uniform matter background, with $n=4$.
    Left panel: $a=0.99M$; right panel: $a=0.2M$.
    }
    \label{beta}
\end{figure}

Figs.~\ref{n}-\ref{beta} further illustrate the properties of the scalar superradiant instability induced by non-uniform matter.
\noindent
The case $\mu_{0}=0$ reproduces the results reported in Ref.~\cite{Cardoso:2013fwa}.
The main features of the instability can be summarized as follows:
\begin{itemize}
    \item 
    The real part of the frequency approximately scales as $1/r_{0}$. 
    For sufficiently small $r_{0}$, the system lies outside the superradiant regime $\omega_{\text{R}} < m\Omega_\text{H}$, and no instability develops.
    \item 
    A smaller radial decay rate of the matter distribution ($n$) allows the superradiant instability to be triggered already at lower values of $\beta$.
    However, for different radial decay rates, the efficiency of the resulting superradiant instability remains comparable, with no significant differences observed.
    \item 
    For fixed $\beta$, increasing $r_{0}$ suppresses and eventually quenches the instability.
    \item 
    Increasing either $\beta$ or $\mu_{0}$ leads to a larger value of $\omega_{\text{R}}$.
    \item 
    At sufficiently large $r_{0}$, and for any sufficiently large $\beta$, the instability persists with an approximately constant growth rate.
\end{itemize}
This behavior closely resembles that of a black hole bomb, in which a reflecting mirror is effectively located at $r_0$ around a Kerr black hole.

The physical origin of these features can be understood from the effective potential analysis.
The radial equation can be cast into a Schrödinger-like form
\begin{align}
    \frac{d^{2} \Psi}{dr_{*}^{2}} + (\omega^{2} - V_{\mathrm{eff}})\Psi =0
    \, ,
\end{align}
where $r_*$ is the tortoise coordinate defined as $dr_*=(r^2+a^2)/\Delta dr$ and $\Psi=\sqrt{r^2+a^2}R_{lm}$. The effective potential is given by
\begin{equation}
V_\text{eff}= \omega^{2}
    - \frac{
    K 
    - \Delta \left[ \mu_{0}^{2} r^{2} +G(r)  +\lambda_{lm} \right]
    }
    {(r^{2} +a^{2})^{2}}+\frac{\Delta}{(r^{2} +a^{2})^{3}} 
    \left(
    r \frac{d\Delta}{dr} + \frac{a^{2}-2r^{2}}{a^{2}+r^{2}} \Delta
    \right).
\end{equation}
\begin{figure}[H]
	\centering
	\includegraphics[width=0.45\linewidth]{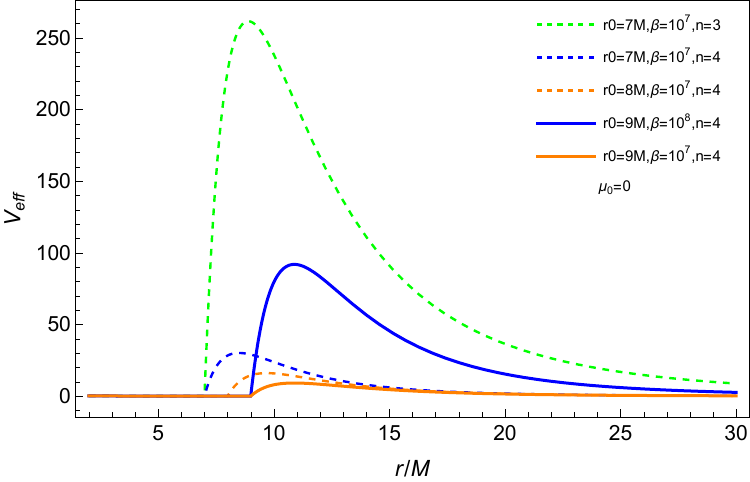}
    \includegraphics[width=0.45\linewidth]{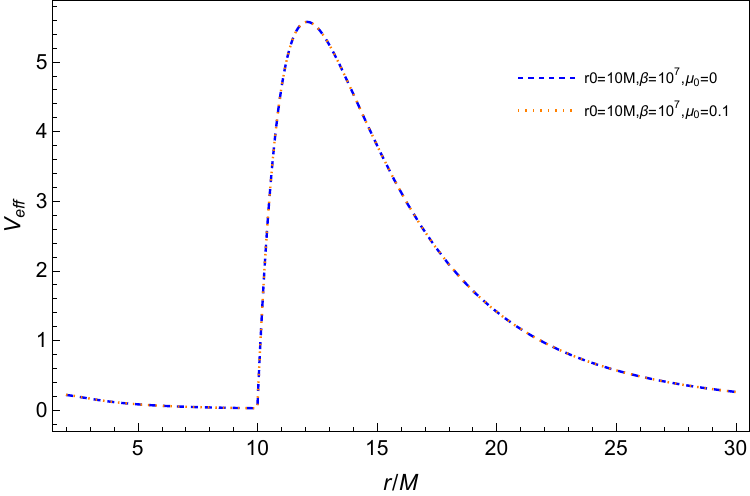}
	\caption{
    (Left) Effective potential for different parameters with $\mu_{0} = 0$ fixed. 
    (Right) Effective potential for the different choices of $\mu_{0}$ with the other parameters fixed: for $\mu_{0}M =0$, $V_{\text{max}} = 6.22323$; and for $\mu_{0}M =0.1$, $V_{\text{max}} = 6.25442$.  
    We fixed the $a=0.99M$ in both plots.
    } 
    \label{V}
\end{figure}
As shown in Fig.~\ref{V}, a slower radial decay of the matter distribution corresponds to a higher effective potential barrier; consequently, the superradiant instability can be triggered at smaller values of $\beta$.
For fixed $\beta$, increasing $r_{0}$ lowers the height of the effective potential barrier, thereby suppressing the superradiant instability. 
For fixed $r_{0}$, a critical value of $\beta$ corresponds to a critical barrier height; once $\beta$ exceeds this threshold, the unstable growth rate approaches an approximately constant value.
By contrast, variations in $\mu_{0}$ have only a minor effect on the barrier height in this setup.

Notably, the shape of the effective potential is influenced more strongly by the environment-induced effective mass than by the intrinsic mass of the scalar field $\mu_0$ itself.
Different matter profiles and coupling strengths, therefore, lead to qualitatively different superradiant behaviors. 
In other words, the chameleon mechanism plays an essential role in determining the superradient dynamics.
As shown in the following subsection, this environmental dependence directly translates into observable differences in the stimulated emission from superradiant scalar clouds.

\subsection{Stimulated radiation}
\label{non-uniform-sr}

\begin{figure}[H]
	\centering          
    \includegraphics[width=0.35\linewidth]{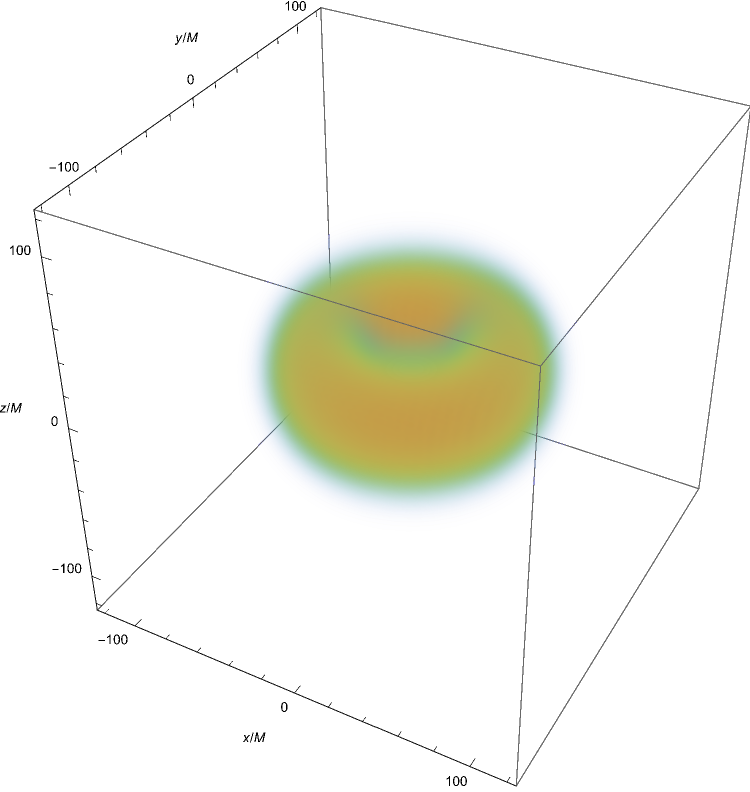}
	\caption{
    In the non-uniform matter, the model of the superradiant “$l=m=1$” scalar cloud around a Kerr black hole.
    }
    \label{non-uniform-sc}
\end{figure}

In this subsection, we investigate stimulated radiation in the presence of a non-uniform matter distribution.
We consider the parameter choice $n=4$, $a = 0.2 M$, $r_{0} = 99.889 M$, $\beta=10^{10}$, $\mu_{0} M = 0.03$, and $\mu_{0}=10^{-5}~\mathrm{eV}$, which leads to
\begin{align}\label{non-uniform-1}
\begin{split}
    M 
    &= 8 \times 10^{23}~\mathrm{kg} 
    \, ,\quad 
    \Gamma_{\varphi} 
    = (3\times10^{40})^{-1}~\mathrm{s}^{-1}
    \, ,\\
    \Gamma_{g} 
    &= 5\times 10^{-2}~\mathrm{s}^{-1}
    \,,\quad
    \Gamma_{e} 
    = 2.5\times 10^{10}~\mathrm{s}^{-1}
    \, ,
\end{split}
\end{align}
with a matter density $\rho_{\text{max}} \approx 4.15 \times 10^{30}~\mathrm{J/m}^{3}$. 
For these parameters, the resulting superradiant cloud exhibits an approximately toroidal geometry, as illustrated in Fig.~\ref{non-uniform-sc}, with characteristic radii $\langle r \rangle = 60 M$ and $\Delta r = 20 M$.
The corresponding volume is given by $V=2\pi^{2}\langle r \rangle \Delta r^{2}$, and we obtain $\beta_{\varphi} \approx 0.16$ and $\Gamma_{e} = 1/\Delta r$.
Under these conditions, curvature effects can be safely neglected in studies of scalar decay.
Consequently, the evolution equations for the scalar and photon numbers are governed by Eq.~\eqref{evolution-eq-l=1}.

We emphasize that for other bosonic fields that do not possess an environment-dependent mechanism, the stimulated emission in a non-uniform matter background coincides with that in a uniform matter distribution. 
On the other hand, the scalar field in scalar-tensor theory exhibits the chameleon mechanism, allowing us to observe differences in the matter distribution.
In other words, the other bosonic fields in uniform and non-uniform matter backgrounds are also equivalent to the scalar field with the chameleon mechanism in the uniform matter background.
As mentioned previously, it is thus essential to investigate the scalar field in the non-uniform matter distribution so that we can distinguish the scalar field with the chameleon mechanism from the other bosonic fields.

\begin{figure}[H]
	\centering
	\includegraphics[width=0.42\linewidth]{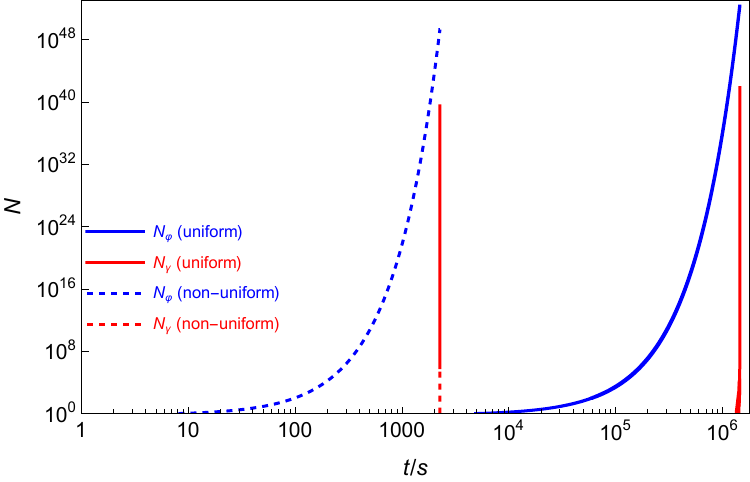}
    \includegraphics[width=0.42\linewidth]{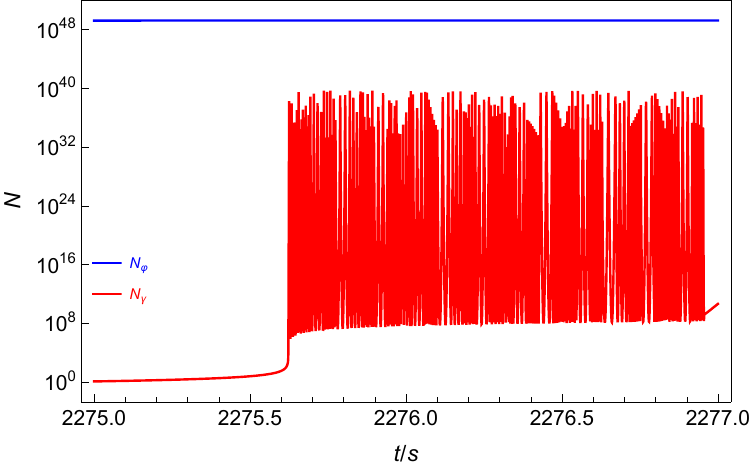}
	\caption{
    Time evolution of the occupation numbers $N_{\varphi}$(blue) and $N_{\gamma}$(red) for the uniform (solid lines) and non-uniform (dashed lines) matter distributions. The results are shown on a log–log scale to cover the wide dynamical range in both time and occupation number.
    The right panel shows an enlarged view of the time when the photon number begins to increase for a non-uniform matter distribution.
    The figure is generated based on the parameter values given in Eq.~\eqref{non-uniform-1} and Eq.~\eqref{uniform-3}.
    }
    \label{l=1-nonuniform-N}
\end{figure}

\begin{figure}[H]
	\centering
	\includegraphics[width=0.42\linewidth]{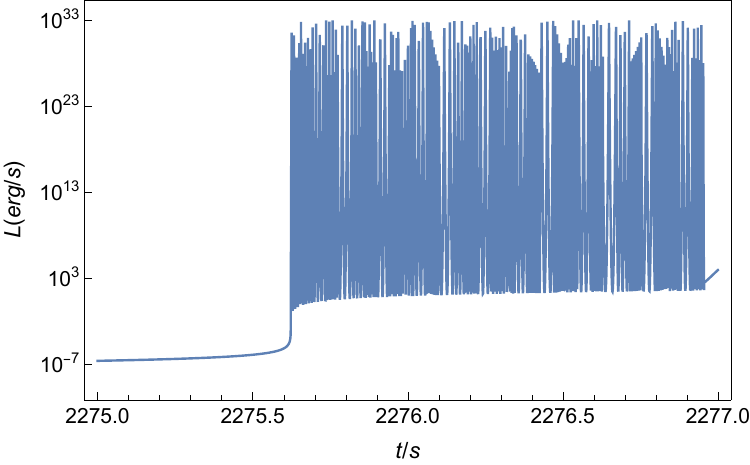}
    \includegraphics[width=0.42\linewidth]{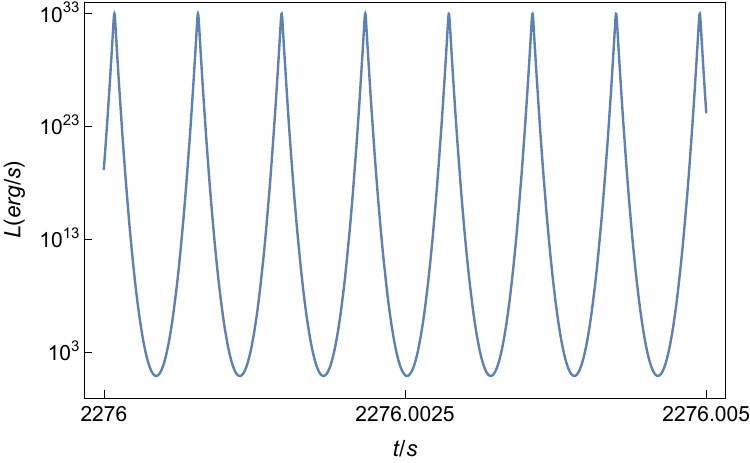}
	\caption{
    Evolution of the average photon luminosity after the stimulated decay takes over in the non-uniform matter case. 
    The parameter $n$ does not affect $\omega_\text{I}$, and we assumed $n=4$.
    The figure is generated based on the parameter values given in Eq.~\eqref{non-uniform-1}.
    }
    \label{l=1-nonuniform-L}
\end{figure}

\begin{figure}[H]
	\centering
	\includegraphics[width=0.42\linewidth]{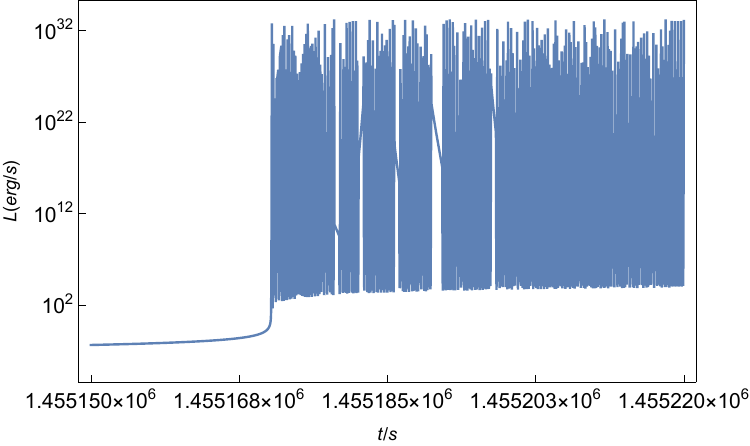}
    \includegraphics[width=0.42\linewidth]{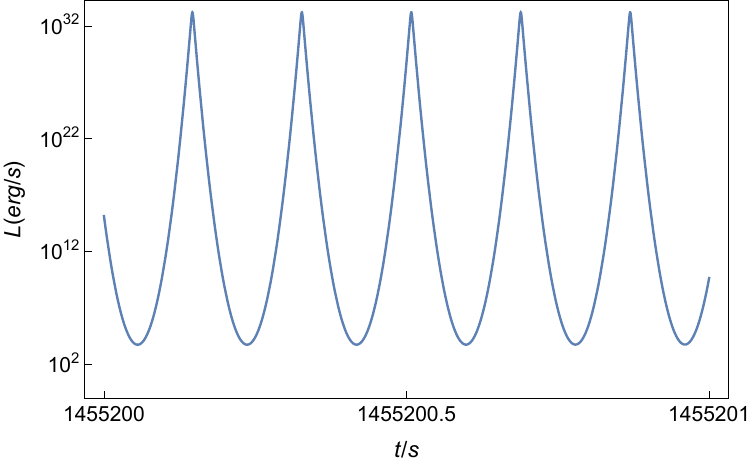}
	\caption{
    Evolution of the average photon luminosity after the stimulated decay takes over in the uniform matter case.
    The figure is generated based on the parameter values given in Eq.~\eqref{uniform-3}.
    }
    \label{l=1-L-0.03}
\end{figure}

In the following, we compare the non-uniform case with the uniform case and discuss the differences in the electromagnetic signals from stimulated emission.
For a new parameter set in common, $a = 0.2 M$, $\mu_{0} M = 0.03$, and $\mu_{0} = 10^{-5}~ \mathrm{eV}$, we obtain
\begin{align}\label{uniform-3}
\begin{split}
    M 
    &= 8 \times 10^{23}~\mathrm{kg} 
    \, ,\quad 
    \Gamma_{\varphi} 
    = (3\times10^{40})^{-1}~\mathrm{s}^{-1}
    \,,\\
    \Gamma_{g} 
    &= 8.3\times 10^{-5}~\mathrm{s}^{-1}
    \,,\quad
    \Gamma_{e} 
    = 2\times 10^{8}~\mathrm{s}^{-1}
    \, .
\end{split}
\end{align}
Definitions of $V$, $\beta_{\varphi}$, and $\Gamma_{e}$ follow those given in the subsection~\ref{uniform-l=1}.
As shown in Fig.~\ref{l=1-nonuniform-N}, the temporal evolution of the scalar and photon numbers in the non-uniform matter case closely resembles that in the uniform matter distribution, and since $\Gamma_{g}$ is larger in non-uniform matter than in uniform matter, the timescale for the onset of stimulated radiation is correspondingly shorter in the non-uniform case.
Since we have adopted $F(\theta) = 0$, the angular dependence of the scalar and photon numbers also remains consistent with that of the uniform case.
However, a clear distinction emerges in the duration of the stimulated radiation.
As can be seen from Figs.~\ref{l=1-nonuniform-L}–\ref{l=1-L-0.03}, the temporal profile of the stimulated emission differs significantly between the non-uniform and uniform matter distributions.
In particular, stimulated radiation from scalars in a non-uniform matter environment exhibits distinct time scales compared to those of other bosonic fields, providing a potential observational handle for discriminating between different models.


To ensure that the superradiant instability persists until the onset of stimulated emission, the angular momentum of the critical scalar cloud must satisfy the condition:
\begin{align}
   J_{\varphi}^{c} = N_{\varphi}^{c} \le J_{\text{BH}} = a M 
   \, ,
\end{align} 
where $N_{\varphi}^{c}$ denotes the critical scalar number required for stimulated emission~\cite{Rosa:2017ury}.
Our choice of parameters and the resulting numerical solutions satisfy this condition.
As the boson cloud grows through superradiance, the particle number $N_{\varphi}$ increases accordingly.
Once it reaches the threshold $N_{\text{bosenova}}$, nonlinear self-interactions become significant and may trigger a collapse of the cloud, known as the bosenova.
To maintain the viability of laser-like emission, such a collapse must be avoided before the scalar cloud reaches the critical number for stimulated emission.

For a uniform matter distribution, the bosenova threshold for scalar particles is approximately given by
\begin{align}
\label{N-bosenova}
    N_{\text{bosenova}} 
    \approx 10^{78} c_{0} 
    \cdot \frac{n'^{4}}{\alpha_{\mu}^{3}} 
    \cdot \left(\frac{M}{M_{\odot}} \right)^{2}
    \cdot \left(\frac{f_{\varphi}}{M_{\mathrm{pl}}} \right)^{2}
    \, ,
\end{align}
where $n'$ is the principal quantum number of the scalar field, $\alpha_{\mu}=\mu_{0} M$, $M_{\odot}$ denotes the solar mass, $f_{\varphi}=1/\beta_{\gamma}~M_{\mathrm{pl}} $ is the decay constant, and $c_{0} \sim 5$ is determined by numerical simulations~\cite{Yoshino:2012kn, Dupuis:2018dhs, Zu:2020whs}.
For our parameter choices, we find $N_{\varphi}^{c} < N_{\text{bosenova}}$.

In the presence of a non-uniform matter distribution, the structure of the superradiant cloud depends sensitively on the matter profile, and the standard estimate in Eq.~\eqref{N-bosenova} is therefore modified.
For the parameters considered in Sec.~\ref{non-uniform-sr}, we obtain $N_{\text{bosenova}} \approx 2 \times 10^{52}$, which remains larger than $N_{\varphi}^{c}$.
This value ensures that the bosenova collapse does not occur before the scalar cloud reaches the critical threshold required for stimulated emission.

\section{conclusions and discussion}\label{sec-cd}

In this work, we have investigated stimulated emission from superradiant scalar clouds generated via the superradiant instability in both uniform and non-uniform matter backgrounds.
We have demonstrated that the stimulated emission produced by such scalar clouds can be sufficiently strong to be detectable, and that the angular distributions of both scalar and photon number densities are determined by the underlying scalar mode.
We have further shown that the environmental dependence of the scalar field leads to qualitatively different superradiant behavior in uniform and non-uniform matter distribution, resulting in potentially distinguishable features in the associated stimulated emission signals.

In the case of uniform matter, the superradiant instability of scalar fields and the resulting stimulated emission from the scalar clouds closely resemble those found for other bosonic fields.
The growth rate of the instability depends on the scalar mode, with the fastest-growing mode being $(l=m=1)$.
For this mode, the stimulated emission is maximized in the equatorial plane and gradually suppressed toward the polar directions.
For the $(l=2, m=1)$ mode, the emission peaks at $\theta=\pi/4$ and is minimized at $\theta=0$ and $\pi/2$.
As a consequence of the angular dependence of the stimulated emission, the scalar distribution within the superradiant cloud evolves toward a spherically symmetric configuration at late times.

In non-uniform matter, due to the chameleon mechanism, the properties of the scalar superradiant instability depend sensitively on the matter distribution and on the coupling strength between the scalar and matter fields.
In such environments, the growth of the superradiant scalar cloud can be more efficient than in the uniform case, while the duration of the resulting stimulated emission is correspondingly shorter.

These results indicate that scalar fields originating from the scalar-tensor theory may leave distinct observational imprints in the form of stimulated emission signals that can be differentiated from those associated with other light bosons.
In particular, the strong dependence on the matter distribution suggests that realistic astrophysical environments, such as accretion disks, could significantly modify both the growth and emission properties of superradiant scalar clouds.
This environmental sensitivity provides a potentially unique observational handle for discriminating scalar-induced signals from those arising from axions or other bosonic degrees of freedom.

Several limitations of the present analysis should be noted.
Our treatment was restricted to idealized matter distributions, whereas realistic astrophysical systems may involve additional effects such as magnetic fields and the dynamical evolution of the surrounding matter.
The matter distribution adopted in this work was motivated by a simplified model of an accretion disk, where the inner edge was described by a step-function–like profile.
Such a sharp transition is not essential, and it is straightforward to replace it with a smoother radial distribution that more realistically captures the gradual variation of the matter density near the inner region of the disk.
This modification would alter the structure of the effective potential in the radial equation, potentially leading to significant changes in the superradiant instability, including its growth rate and the parameter range in which it occurs.

Furthermore, the coupling strength between the scalar and matter fields remains uncertain, and this parameter plays a crucial role in determining both the efficiency of the superradiant instability and the amplitude of the observable signals.
In this respect, complementary constraints from ground-based searches for the light scalar fields may play an important role in narrowing the viable parameter space.
A combined analysis of astrophysical observations and ground-based experiments would therefore provide a powerful approach to probing the properties of scalar fields arising from the scalar-tensor theory.
Future work should aim to incorporate more realistic astrophysical environments and to develop quantitative predictions for observational signatures in specific frequency bands.

\section*{Acknowledgement}

T.K. is supported by National Key R\&D Program of China (No.~2021YFA0718500) and the National Natural Science Foundation of China (No.~12403003).
S.N. is supported by JSPS KAKENHI Grant No.~24K17053.


\bibliographystyle{apsrev4-1}
\bibliography{References}

\end{document}